\documentclass[11pt]{article}
\usepackage{amsfonts
}
\usepackage{amsmath,float}
\usepackage{amssymb}
\usepackage{graphicx}
\usepackage{color}
\usepackage{braket}
\usepackage[utf8]{inputenc}
\usepackage{hyperref}
\pdfoutput=1

\def \be {\begin{equation}}
\def \ee {\end{equation}}
\def \ba {\begin{aligned}}
\def \ea {\end{aligned}}
\def \bea {\begin{eqnarray}}
\def \eea {\end{eqnarray}}

\begin{document}

\begin{titlepage}
\begin{flushright}
\end{flushright}
\vspace{0.5cm}
\begin{center}
{\Large \bf TBA-like equations  for non-planar scattering amplitude/Wilson lines duality at strong coupling
}
\lineskip .75em
\vskip 2.5cm
{Hao Ouyang$^{a,}$\footnote{haoouyang@jlu.edu.cn}
and Hongfei Shu $^{b,c,}$\footnote{shuphy124@gmail.com }
}
\vskip 2.5em
 {\normalsize\it $^{a}$Center for Theoretical Physics and College of Physics, Jilin University, Changchun 130012, China\\
$^{b}$Beijing Institute of Mathematical Sciences and Applications (BIMSA), Beijing, 101408, China\\
$^{c}$Yau Mathematical Sciences Center (YMSC), Tsinghua University, Beijing, 100084, China

}
\vskip 3.0em
\end{center}
\begin{abstract}
We compute the minimal area of a string worldsheet ending on two  infinite periodic light-like Wilson lines in the AdS$_3$ boundary, which is dual to the first non-planar correction to the gluon scattering amplitude in $\mathcal{N}=4$ SYM at the strong coupling. Using the connection between the Hitchin system and the thermodynamic Bethe ansatz (TBA) equations, we present  an analytic method to compute the minimal area surface and express the non-trivial part of the minimal area in terms of the free energy of the TBA-like equations. Given the cross ratios as inputs, the area computed from the TBA-like equations matches that calculated using the numerical integration. 
  \end{abstract}
\end{titlepage}

\tableofcontents

\section{Introduction}
The AdS/CFT correspondence provides a powerful method to study the non-perturbative gauge theory \cite{Maldacena:1997re}. Especially, the integrable features in the four dimension maximally supersymmetric Yang-Mills theory have led to many novel results in the past decades \cite{Beisert:2010jr}. One of the most significant achievements is the scattering amplitude/Wilson loop duality \cite{Alday:2007hr}. On the AdS side, the gluon scattering amplitude is mapped to a worldsheet amplitude ending on an IR D3 brane near the AdS horizon. Performing T-duality transformations and redefining the radial coordinate, one obtains a worldsheet ending on a light-like Wilson loop in the T-dual AdS boundary \cite{Alday:2007hr}, whose minimal area provides the amplitude at the strong coupling \cite{Maldacena:1998im,Rey:1998ik}. The equations of motion and the Virasoro constraints, which determine the minimal area surface in AdS$_3$ (resp. AdS$_5$), are reduced to the classically integrable equations \cite{Pohlmeyer:1975nb} or equivalently to the $SU(2)$ (resp. $SU(4)$) Hitchin system \cite{Alday:2009yn,Alday:2009dv,Burrington:2009bh,Burrington:2011eh} \footnote{We will denote it by linear problem instead of Hitchin system in the main text.}.

The Stokes data of the $SU(2)$ Hitchin system has been well studied in a very different context, i.e. the wall-crossing of the BPS spectrum in ${\cal N}=2$ Super Yang-Mill theory \cite{GaiottoMooreNeitzke:2008,GaiottoMooreNeitzke:2009}, where a connection between the Hitchin system and the thermodynamic Bethe ansatz (TBA) equations has been found \footnote{This connection is now known as the ODE/IM correspondence \cite{Dorey:1998pt,Bazhanov:1998wj,Lukyanov:2010rn}. }. To study the minimal surface with a light-like polygonal boundary condition, an irregular singularity should be imposed in the linear problem, whose Stokes data was used to construct the boundary. Inspired by the connection between the Hitchin system/linear problem and TBA equations, one finds that the non-trivial part of the minimal area can be expressed by the free energy of the TBA equations and the Y-system \cite{Alday:2010vh,Hatsuda:2010cc,Hatsuda:2010vr}\footnote{Based on the similarity of the Riemann-Hilbert problem, the ODE/IM correspondence for the Schr\"odinger equation with arbitrary polynomial potentials has been studied in \cite{Ito:2018eon,Ito:2019jio,Emery:2020qqu}.}. See also \cite{Fioravanti:2020udo} from the approach of the QQ-system and the non-linear integral equations (NLIEs). 

The scattering amplitude has been generalized to the form factor \cite{Alday:2007he}, whose operator corresponds to a closed string extending from the original AdS boundary and inserted on the scattering worldsheet.
After the T-duality transformations, the boundary becomes a periodic light-like Wilson line, whose period is determined by the momentum of the operator.
The worldsheet is furthermore extended to the T-dual AdS horizon. The problem of the form factor reduces to the one of computing the minimal area ending on one period, which is also encoded in the free energy of the TBA system \cite{Maldacena2010, Gao:2013dza}. 

The non-planar scattering amplitude in the context of AdS/CFT correspondence was not well explored for a long time. The main difficulty is due to the higher genus of the Riemann surface of the non-planar case. One beautiful idea to overcome this difficulty is to cut the higher genus Riemann surface into disks, where the planar techniques can be applied, and then glue them together. Based on this idea the first non-planar correction of the scattering amplitudes/Wilson loop duality was first proposed by Ben-Israel, Tumanov and Sever in \cite{Ben-Israel:2018ckc}.

Let us consider the $1/N$ correction of scattering amplitude, i.e. a double trace amplitude, where $\hat{n}$ gluons in one trace and $\hat{m}$ gluons in the other, which we will denote by $A_{\hat{n},\hat{m}}$. The string dual of $A_{\hat{n},\hat{m}}$ has a topology of cylinder, whose two boundaries end on the IR D3 brane. The two traces correspond to two boundaries of the cylinder. On one boundary, $\hat{n}$ vertex operators are inserted, whose momentum are denoted by $(k_1, k_2, \cdots, k_{\hat{n}})$. The other $\hat{m}$ vertex operators with momentum $(k_{\hat{n}+1}, k_{\hat{n}+2},\cdots, k_{\hat{n}+\hat{m}})$ are inserted on the other boundary. The total momentum of each boundary is given by
\be
q=\sum_{i=1}^{\hat{n}}k_i=-\sum_{j=1}^{\hat{m}}k_{\hat{n}+j}.
\ee 
To apply the planar techniques, one cuts the cylinder into a disk \cite{Ben-Israel:2018ckc}. The cut (curve $\gamma$) starts from one boundary of the cylinder and ends at the other boundary. The curve $\gamma$ on the diagram crosses a certain number of propagators. Set $P_{\gamma(j)}$ as the momentum that crosses the cut in the direction coinciding with the external particle ordering $(1, 2, \cdots, \hat{n})$.
$l=\sum_jP_{\gamma(j)}$ is interpreted as the momentum flow around the cylinder. Then the full amplitude $A_{\hat{n}, \hat{m}}$ is given by integration of cut amplitude with respect to $l$:
\be
A_{\hat{n},\hat{m}}=\frac{\lambda}{N}\int d^{4}l{\cal A}_{\hat{n},\hat{m}}^{\gamma}(l).
\ee
One could start with a curve $\gamma'$ winding around the cylinder once than $\gamma$, then the momentum $l$ is shifted by the total momentum $q$: ${\cal A}_{\hat{n},\hat{m}}^{\gamma^{\prime}}(l)={\cal A}_{\hat{n},\hat{m}}^{\gamma}(l+q)$.
Then $l$ is only well defined modulo a shift by the total momentum $q$: $l\simeq l+q$. To construct an unambiguously defined quantity, one has to sum over all possible shifts of $l$ by the integer number of $q$, i.e.
\be
\mathbb{A}_{\hat{n},\hat{m}}(l)=\sum_{a}{\cal A}_{\hat{n},\hat{m}}^\gamma (l+aq),\quad {A}_{\hat{n},\hat{m}}=\frac{\lambda}{N}\int_{l\simeq l+q}\mathbb{A}_{\hat{n},\hat{m}}(l).
\ee

Performing the T-dual transformation on the four directions of the IR D3 brane and redefining the radial coordinate, one obtains an AdS$_5\times S^5$ spacetime again. Under the T-dual transformation, the worldsheet action of the amplitude becomes a Polyakov action of T-dual $y$ coordinates with the periodic condition
\be
y(\tau,\sigma=\gamma(\tau)+2\pi)=y(\tau,\sigma=\gamma(\tau))+q
\ee
and condition the AdS boundary:
\be
\ba
y(\tau=0,\sigma_i<\sigma<\sigma_{i+1})&=-\sum_{p\leq i}k_p-c\\
y(\tau=L,\sigma_{j}<\sigma<\sigma_{j+1})&=\sum_{p\leq j}k_{\hat{n}+p}-c+l,
\ea
\ee
where $\tau \in [0,L]$ and $\sigma \in S^1$ are the worldsheet coordinates. $\sigma_i$ is the insertion point of the vertex operator. $c$ is an arbitrary constant. One thus obtains two Wilson lines with $\hat{m}$ and $\hat{n}$ segments, respectively. Since the gluons are massless, these segments are light-like. This boundary condition implies that the worldsheet after T-dual transforms ends on two periodic light-like Wilson lines on the T-dual AdS boundary. This thus generalizes the scattering amplitude/Wilson loop duality to the non-planar case \cite{Ben-Israel:2018ckc}:
\begin{description}
\item[Double trace scattering amplitude/Periodic Wilson lines]: the cut double trace scattering amplitude $\mathbb{A}_{\hat{n},\hat{m}}(l)$ is dual to the correlation function of two periodic light-like polygonal Wilson lines.
\end{description}
The duality was also tested perturbatively at one-loop in the SYM side in \cite{Ben-Israel:2018ckc}. At the strong coupling, the amplitude $\mathbb{A}_{\hat{n},\hat{m}}(l)$ can be computed from the minimal area of the worldsheet ending on the Wilson lines. However, the minimal area is not studied in a long time, because of the complicated boundary condition of the worldsheet.
In this paper, we propose a boundary condition of the linear problem to produce the two light-like polygonal Wilson lines at the boundary. We then present a method to exactly compute the minimal area of the worldsheet ending on the Wilson lines with fixed $q$ and $l$ \footnote{Our approach to compute minimal area is inspired by \cite{Janik:2011bd,Kazama:2011cp,Caetano:2012ac}, which compute the correlation function of heavy operators in the AdS part. In their case, the worldsheet is a sphere with punctures, whereas in our case the worldsheet has the topology of cylinder/disk. }. For simplicity, we will focus on the AdS$_3$ spacetime, where only the Wilson lines with even segments are possible, say $\mathbb{A}_{2n,2m}(l)$.

This paper is organized as follows. In section \ref{sec:2}, we first recall the Pohlmeyer reduction of the equation of motion and the calculation of the minimal area in the AdS$_3$ spacetime. We then propose the boundary condition of the linear problem, which produces the minimal surface ending on the light-like Wilson lines at the AdS$_3$ boundary. In section \ref{sec:wkb}, we study the WKB approximation of the linear problem to extract the data that is needed in the calculation of the minimal area. We introduce the Fock-Goncharov coordinates associated with the linear problem, which correspond to the cross ratios of the Wilson lines. By using the TBA-like equations satisfied by the Fock-Goncharov coordinates, we express the minimal area in an analytic form. In section \ref{sec:phy-con}, we compute the minimal area from TBA-like equations with the physical cross ratios. We also test our method by comparing it with the numerical integration of the area. The section \ref{sec:con-dis} is devoted to conclusions and discussion. In appendix \ref{sec:m=n}, we present simplified functional relations and TBA equations for the case $m=n$, where the connection with the ${\cal N}=2$ super Yang-Mills theory is also mentioned.

\section{Classical string in AdS$_3$ and minimal area}\label{sec:2}
In this section, we first recall the Pohlmeyer reduction of the equation of motion and the Virasoro constraints of the classical strings in the AdS$_3$ spacetime, and then propose a boundary condition of the linear problem, which leads to the minimal surface ending on two light-like Wilson lines at the AdS$_3$ boundary. Based on the boundary condition of the linear problem, we introduce small solutions of the linear problem, whose combination is used to express the cross ratios of the Wilson lines. We finally show the calculation of the minimal area from the solutions of the generalized sinh-Gordon equation. 

\subsection{Pohlmeyer reduction and the linear problem}
The AdS$_3$ spacetime can be written as a surface embedding in $\mathbb{R}^{2,2}$ with the constraint
\be
\vec{Y}\cdot \vec{Y}=-Y_{-1}^{2}-Y_{0}^{2}+Y_{1}^{2}+Y_{4}^{2}=-1.
\ee
Classical strings in AdS$_3$ are described by the equation of motion and the Virasoro constraints
\be\label{eq:eom-Vir}
\partial\bar{\partial}\vec{Y}-(\partial\vec{Y}\cdot\bar{\partial}\vec{Y})=0,\quad
\partial\vec{Y}\cdot\partial\vec{Y}=0=\bar{\partial}\vec{Y}\cdot\bar{\partial}\vec{Y},
\ee
which are equivalent to the generalized sinh-Gordon equation 
\be
\label{mshG}
\partial_z\partial_{\bar{z}}\alpha-e^{2\alpha}+p(z)\bar{p}(\bar{z})e^{-2\alpha}=0
\ee
according to the Pohlmeyer reduction \cite{Pohlmeyer:1975nb,Alday:2009yn}. Here $\alpha$ and $p$ are $SO(2,2)$ invariant function:
\be
\ba
\label{eq:p-fun}
e^{2\alpha}&=\frac{1}{2}\partial\vec{Y}\cdot \bar\partial\vec{Y},\quad  N_a=\frac{1}{2}e^{-2\alpha}\epsilon_{abcd}Y^b\partial Y^c \bar{\partial}Y^d,\\
\quad
p(z)&=-\frac{\vec{N}}{2}\cdot \partial^2\vec{Y},\quad \bar{p}(\bar{z})=\frac{\vec{N}}{2}\cdot \bar{\partial}^2\vec{Y}.
\ea
\ee
The equation of motion and Virasoro constraints \eqref{eq:eom-Vir} are equivalent to the linear problem
\be			\partial_{z}\psi+B_{z}\psi=0, \quad \partial_{\bar{z}}\psi+B_{\bar{z}}\psi=0
\ee
with the connections\footnote{The flatness condition of this linear problem can be rephrased as $D_{\bar{z}}\Phi_z=D_z\Phi_{\bar{z}}=0$ and $F_{z\bar{z}}+[\Phi_z,\Phi_{\bar{z}}]$, which are the Hitchin system \cite{Hitchin:1986vp}.  $A$ and $\Phi$ have the interpretation of the gauge connection and Higgs field, respectively, in two dimensions. 
}:

\be
\ba\label{eq:connection}
B_{z}&=\left(\begin{array}{cc}				\frac{1}{2}\partial_{z}\alpha & -\frac{1}{\zeta}e^{\alpha}\\
	-\frac{1}{\zeta}e^{-\alpha}p(z) & -\frac{1}{2}\partial_{z}\alpha
	\end{array}\right)=:A_z+\Phi_z,\\
		B_{\bar{z}}&=\left(\begin{array}{cc}
		-\frac{1}{2}\partial_{\bar{z}}\alpha & -\zeta e^{-\alpha}\bar{p}(\bar{z})\\
				-\zeta e^{\alpha} & \frac{1}{2}\partial_{\bar{z}}\alpha
		\end{array}\right)=:A_{\bar{z}}+\Phi_{\bar{z}},
\ea
\ee
where $\zeta$ is a complex value called the spectral parameter. The flatness condition of the connections with any complex value $\zeta$ leads to the generalized sinh-Gordon equation \eqref{mshG}. Solving the linear problem at $\zeta=1$ and $\zeta=i$, one can construct the AdS$_3$ coordinates \cite{Alday:2009yn}:
\be
\label{st-cd}
\left(\begin{array}{cc}
Y_{-1}+Y_{4} & Y_{1}-Y_{0}\\
Y_{1}+Y_{0} & Y_{-1}-Y_{4}
\end{array}\right)_{a,\dot{a}}=\psi_{\alpha,a}^LM_{\alpha\dot{\beta}}\psi_{\dot{\beta},\dot{a}}^R,
\ee
where $M$ is a matrix depending on the gauge. $\psi_{\alpha,a}^L$ and $\psi_{\alpha,a}^R$ are the solutions of the linear problem with $\zeta=1$ and $\zeta=i$, respectively.

\subsection{Boundary condition of the linear problem}
To study the minimal surface with a polygonal-type boundary, it is convenient to take advantage of the linear problem. The minimal surface associated with the scattering amplitude/light-like polygon Wilson loop are characterized by a polynomial $p(z)$ and boundary condition of $\alpha$, $e^{2\alpha }\sim \sqrt{p\bar{p}}$ at $z\to \infty$. The linear problem in this case has an irregular singular point at $z\to \infty$, whose Stokes phenomena leads to the null polygonal boundary condition at AdS boundary \cite{Alday:2009yn}.

To produce two light-like polygonal Wilson lines at AdS boundary, we impose two irregular singular points in $p(z)$ with boundary conditions: $\hat{\alpha}=\alpha-\frac{1}{4}\log(p\bar{p})$ vanish at the irregular singular point and  is regular anywhere on the worldsheet\footnote{As shown in section \ref{sec:phy-con}, we will relax this condition at the zeros of $p(z)$ when the cross ratios ${\cal X}(\zeta=1, i)$ are negative.}. A natural choice of $p(z)$ describing the two light-like polygonal Wilson lines with $2m$ and $2n$ segments respectively is 
\be
p(z)=z^{n-2}+\cdots+\frac{1}{z^{m+2}},
\ee
where the irregular singular points are located at $z=0, \infty$. 

Since the growing solution of the linear problem at $z\to 0, \infty$ will lead to some divergent components in the string coordinates (\ref{st-cd}), one thus expects the worldsheet attaches the AdS boundary when $z\to 0,\infty$. There thus will be two Wilson lines, say $(i)=(0), (\infty)$, at AdS boundary. The solution $\psi$ can be expressed by using the ``big solution'' $b$ and the ``small solution'' $s$ of each sector:
\be
\psi_a=c_a^b b+c_a^s s,
\ee
where only the big solution dominants in the calculation of string coordinates. To extract the information of the big solution, i.e. the coefficient $c_a^b$, we take the product of $\psi_a$ and small solution $s$. One thus can express the AdS coordinates by
\be
Y_{a\dot{a}}=(\psi_{a}^{L}\wedge s^{L})(\psi_{\dot{a}}^{R}\wedge s^{R})(b_{\alpha}^{L}M^{\alpha\dot{\beta}}b_{\dot{\beta}}^{R}),
\ee
which implies $\vec{Y}^2=0$, i.e. the AdS boundary. It is useful to introduce the light-cone coordinates
\be
	(y_{k}^{(i)})^{\pm}=\big(\frac{Y_{1}\pm Y_{0}}{Y_{-1}+Y_{4}}\big)_{k}^{(i)}=\frac{\psi_{2}^{L,R}\wedge(s_{k}^{(i)})^{L,R}}{\psi_{1}^{L,R}\wedge(s_{k}^{(i)})^{L,R}},
\ee
where $k$ is the label of the cusp along the $(i)$-Wilson line. We are then able to compute the distance $(y_{k_{1}k_{2}}^{(i_{1},i_{2})})^{\pm}=(y_{k_{1}}^{(i_{1})})^{\pm}-(y_{k_{2}}^{(i_{2})})^{\pm}$ 
\be
(y_{k_{1}k_{2}}^{(i_{1},i_{2})})^{\pm}=-\frac{\psi_{1}^{L,R}\wedge\psi_{2}^{L,R}(s_{k_{1}}^{(i_{1})})^{L,R}\wedge(s_{k_{2}}^{(i_{2})})^{L,R}}{\psi_{1}^{L,R}\wedge(s_{k_{1}}^{(i_{1})})^{L,R}\psi_{1}^{L,R}\wedge(s_{k_{2}}^{(i_{2})})^{L,R}},
\ee
from which we obtain the cross ration $\chi_{k_1k_2k_3k_4}^{(i_1,i_2,i_3,i_4)\pm}$:
\be\label{eq:cross}
\ba
\chi_{k_1k_2k_3k_4}^{(i_1,i_2,i_3,i_4)\pm}:=\frac{(y_{k_{1}k_{2}}^{(i_{1,}i_{2})})^{\pm}(y_{k_{3}k_{4}}^{(i_{3,}i_{4})})^{\pm}}{(y_{k_{1}k_{3}}^{(i_{1,}i_{3})})^{\pm}(y_{k_{2}k_{4}}^{(i_{2,}i_{4})})^{\pm}}=\frac{(s_{k_{1}}^{(i_{1})})^{L,R}\wedge(s_{k_{2}}^{(i_{2})})^{L,R}(s_{k_{3}}^{(i_{3})})^{L,R}\wedge(s_{k_{4}}^{(i_{4})})^{L,R}}{(s_{k_{1}}^{(i_{1})})^{L,R}\wedge(s_{k_{3}}^{(i_{3})})^{L,R}(s_{k_{2}}^{(i_{2})})^{L,R}\wedge(s_{k_{4}}^{(i_{4})})^{L,R}}.
\ea
\ee
Therefore, the small solutions of the linear problem are important to write down the cross ratios, which will be our main task in the next subsection.

\subsection{Small solution}
At $z=0,\infty$, we are able to diagonalize $\Phi(z)$ and $\Psi_{\bar{z}}$ in the connections, from which we determine the basis of the solutions of the linear problem:
\be
\ba
\psi_{a}\sim&
\exp\Big((-1)^{a}\frac{1}{\zeta}\int^{z}\sqrt{p(z^{\prime})}dz^{\prime}+(-1)^{a}\zeta\int^{\bar{z}}\sqrt{\bar{p}(\bar{z}^{\prime})}d\bar{z}^{\prime}\Big),
\ea
\ee
for $z\to 0, \infty$. Since $z=0,\infty$ are irregular singular points of the linear problem, the complex plane around each sector divides into several sectors, i.e. Stokes sectors, due to the Stokes phenomena. 
It is convenient to introduce a new coordinate $w$ by
\be
dw=\sqrt{p}dz.
\ee
Then at large $w$, the solution to the linear problem is approximated by\footnote{Here we have chosen the gauge to simplify the problem, which does not affect the discussion about the Stokes sectors.}
\be
\psi\sim c_1 e^{\frac{1}{\zeta}w+\zeta\bar{w}}\left(\begin{array}{c}
1\\
0
\end{array}\right)+c_2 e^{-\frac{1}{\zeta}w-\zeta\bar{w}}\left(\begin{array}{c}
0\\
1
\end{array}\right).
\ee
The stokes sector $\hat{{\cal S}}(\zeta)$ on the $w$-plane is given by
\be
\hat{{\cal S}}_{j}(\zeta):(j-\frac{1}{2})\pi+\arg(\zeta)<\arg(w)<(j+\frac{1}{2})\pi+\arg(\zeta).
\ee
At $z\to \infty$, the sectors become
\be\label{eq:ssectorinf}
{\cal S}_{j}^{(\infty)}(\zeta):(j-\frac{1}{2})\frac{2}{n}\pi+\frac{2}{n}\arg(\zeta)<\arg(z)<(j+\frac{1}{2})\frac{2}{n}\pi+\frac{2}{n}\arg(\zeta).
\ee
Therefore, there are $n$ sectors at $z\to \infty$. At $z\to 0$, one finds
\be\label{eq:ssector0}
{\cal S}_{j}^{(0)}(\zeta):(-j-\frac{1}{2})\frac{2}{m}\pi-\frac{2}{m}\arg(\zeta)<\arg(z)<(-j+\frac{1}{2})\frac{2}{m}\pi-\frac{2}{m}\arg(\zeta),
\ee
which leads to $m$ sectors. In each sector, only the decaying solution is uniquely defined. We call these decaying solutions as small solutions.
Let us denote the small solution in sector ${\cal S}^{(i)}_k$ by $s^{(i)}_k$, where $(i)=(0), (\infty)$.

The connection is invariant under the $\mathbb{Z}_2$ projection
\be
\sigma_3B_{z,\bar{z}}(\zeta)\sigma_3=B_{z,\bar{z}}(e^{i\pi}\zeta),
\ee
which enables us to generate the solution of the linear problem by
\begin{eqnarray}
	s_{k+1}^{(i)}=(i\sigma^3)^{k}s_1^{(i)}(e^{ki\pi}\zeta).
\end{eqnarray}
We normalize the small solutions such that
\be
s^{(i)}_k\wedge s^{(i)}_{k+1}=1,
\ee
where the product is defined by $s_a\wedge s_b=\det(s_a,s_b)$.

\subsection{Minimal area}
At the end of this section, let us show how to compute the minimal area from the solutions of the generalized sinh-Gordon equation \eqref{mshG}.
The minimal area ending on the Wilson lines at strong coupling is computed by
\be
A=2 \int d^{2}z \partial\vec{Y}\cdot \bar\partial\vec{Y}=4\int d^{2}ze^{2\alpha},
\ee
where $\alpha$ satisfies the generalized sinh-Gordon equation with the given boundary condition. Since $e^{2\alpha}\sim\sqrt{p\bar{p}}$ at $|z|\to 0,\infty$, which is divergent, we separate the area by
\be
A =4\int d^{2}z(e^{2\alpha}-\sqrt{p\bar{p}})+4\int d^{2}z\sqrt{p\bar{p}},
\ee
and denote the finite part and the divergent part by
\be
A_{{\rm fin}}=4\int d^{2}z(e^{2\alpha}-\sqrt{p\bar{p}}),\quad A_{{\rm div}}=4\int d^{2}z\sqrt{p\bar{p}}.
\ee

The divergent part $A_{\rm div}$ can be regularized by introducing two cutoffs in the radial direction of AdS spacetime:
\be
\ba
A_{\rm div}&=A_{\rm period}+A_{\rm cutoff},\\
A_{\rm period}&=4\int d^{2}z\sqrt{p\bar{p}}-4\int_{\Sigma} d^{2}z\sqrt{p\bar{p}},\\
A_{\rm cutoff}&=4\int_{\Sigma_0, r<\epsilon_0, r<\epsilon_\infty} d^{2}z\sqrt{p\bar{p}}
=A_{\rm cutoff,\epsilon_0}+A_{\rm cutoff,\epsilon_\infty},
\ea
\ee
where $\Sigma$ is a reference surface and $r$ is the radial coordinate with the metric $ds^2=(dy^+dy^-+dr^2)/r^2$. $A_{\rm period}$ depends on the branch cuts and can be evaluated by using the Riemann bilinear identity. $A_{\rm cutoff}$ involves the large $|z|$ and small $|z|$ regions of the Riemann surface, which depends on the cutoff $\epsilon_0$ and $\epsilon_\infty$.
Here we suppose the physical cutoff, i.e. the cutoff on the $r$-direction, $\epsilon_0$ and $\epsilon_\infty$ are small, while the corresponding worldsheet coordinates $1/|z(\epsilon_0)|$ and $|z(\epsilon_\infty)|$ are large.

Since our $p(z)$ and the boundary condition of worldsheet have the same forms as the ones in scattering amplitude case at large/small $|z|$,  $A_{\rm cutoff, \epsilon_0}$ and $A_{\rm cutoff,\epsilon_\infty}$ have the same forms as the ones studied in \cite{Alday:2009dv}
\be
A_{{\rm cutoff},\epsilon_0}=\frac{1}{8}\sum_{i}\big(\log(\epsilon^{2}_0 d_{i,i+2}^{(0)})\big)^{2}+A_{{\rm BDS-like}}^{(0)},
\ee
and
\be
A_{{\rm cutoff},\epsilon_\infty}=\frac{1}{8}\sum_{i}\big(\log(\epsilon^{2}_\infty d_{i,i+2}^{(\infty)})\big)^{2}+A_{{\rm BDS-like}}^{(\infty)},
\ee
where the first term is the usual divergent term, the second term is finite whose detail form can be found in \cite{Alday:2009yn,Alday:2009dv,Alday:2010vh}.

Using the equation of motion (\ref{mshG}), one can rewrite $A_{\rm fin}$ by
\be
A_{{\rm fin}}=2\int d^{2}z\big(e^{2\alpha}-2\sqrt{p\bar{p}}+p\bar{p}e^{-2\alpha}+\partial\bar{\partial}\alpha\big).
\ee
The integral of $\partial\bar{\partial}\alpha$ contributes at boundaries at $|z|\to 0$ and $|z|\to \infty$. At $|z|\to 0$, one finds
\be
2\int d^{2}z\partial\bar{\partial}\alpha \sim\frac{1}{4}2\int d^{2}z\partial\bar{\partial}\log p\bar{p}\sim
\begin{cases}
 \frac{m+2}{2}\int d^{2}z\partial\bar{\partial}\log z\bar{z}  = \frac{\pi}{2}(m+2)&
|z|\to 0\\
\frac{n-2}{2}\int d^{2}z\partial\bar{\partial}\log z\bar{z}=\frac{\pi}{2}(n-2) & |z|\to\infty
\end{cases}.
\ee
Let us denote the first three terms in $A_{\rm fin}$ by $2A_{\rm reg}$:
\begin{align}
A_{\rm fin}&=2A_{\rm reg}+\frac{\pi}{2}(m+n),\label{eq:afin}\\
 A_{{\rm reg}}&=\int d^{2}z\big(e^{2\alpha}-2\sqrt{p\bar{p}}+p\bar{p}e^{-2\alpha}\big).   \label{Areg-org}
\end{align}
The integrand of $A_{\rm reg}$ can be written by
\be
\sqrt{p\bar{p}}e^{2\alpha-\log\sqrt{p\bar{p}}}+\sqrt{p\bar{p}}e^{-2\alpha+\log\sqrt{p\bar{p}}}-2\sqrt{p\bar{p}}=\lambda u,
\ee
with 
\be
\lambda=\sqrt{p},\quad u=2\sqrt{\bar{p}}\big(\cosh(2\hat{\alpha})-1\big).
\ee
To construct a closed form in the integral, we add $vdz=\frac{1}{\sqrt{p}}(\partial\hat{\alpha})^{2} dz$ to one-form $udz$:
\be
\label{Areg}
A_{{\rm reg}}=\frac{i}{2}\int \lambda dz\wedge \eta,
\ee
where $\eta=(ud\bar{z}+vdz)$. By using the Riemann bilinear identity, this integral reduces to integrals over cycles on the double cover of the worldsheet. In the following of this paper, we will provide a procedure to compute this $A_{\rm reg}$ exactly.

\section{WKB approximation and TBA-like equations}\label{sec:wkb}

When $\zeta\to 0,\infty$, we can solve the linear problem by using WKB approximation \cite{GaiottoMooreNeitzke:2009}, where the information needed in the calculation of minimal area, i.e. $\lambda$ and $\eta$ along certain paths, are included. In this section, we first show the WKB approximation of the linear problem by
following the Gaiotto-Moore-Neitzke formalism \cite{GaiottoMooreNeitzke:2009, Caetano:2012ac}. We then introduce the Fock-Goncharov coordinates, which are the cross ratios of the small solutions, and derive their functional relations and integral equations. These integral equations have the form of the TBA equations and enable us to extract the data in the calculation of minimal area. We finally test our method by comparing the area computed from the TBA equations with the one obtained from the numerical integration of \eqref{Areg-org}.

\subsection{WKB curve and WKB triangulation}
\begin{figure}[htb]
\begin{center}
\resizebox{45mm}{!}
{\includegraphics{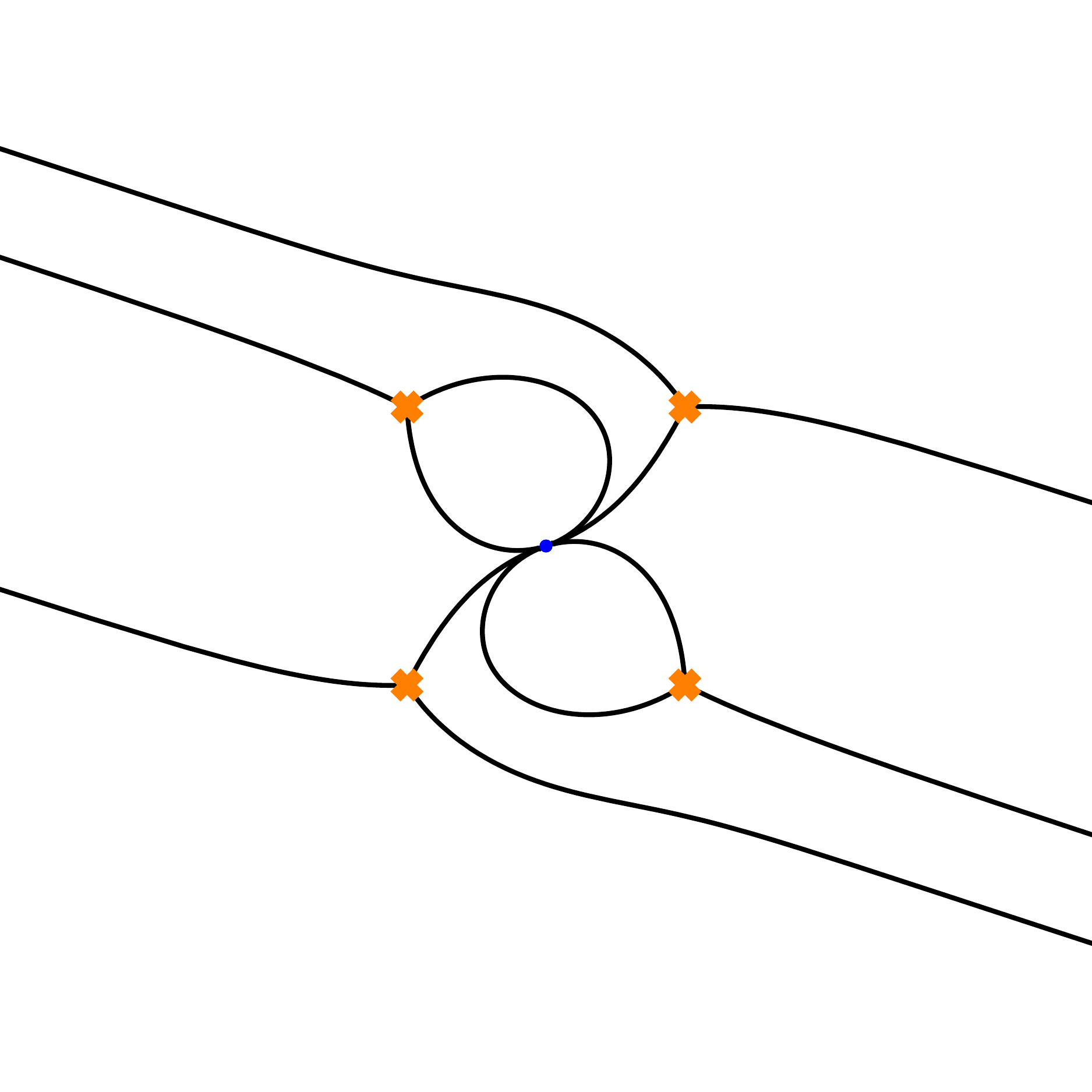}}~~~~
\resizebox{45mm}{!}{\includegraphics{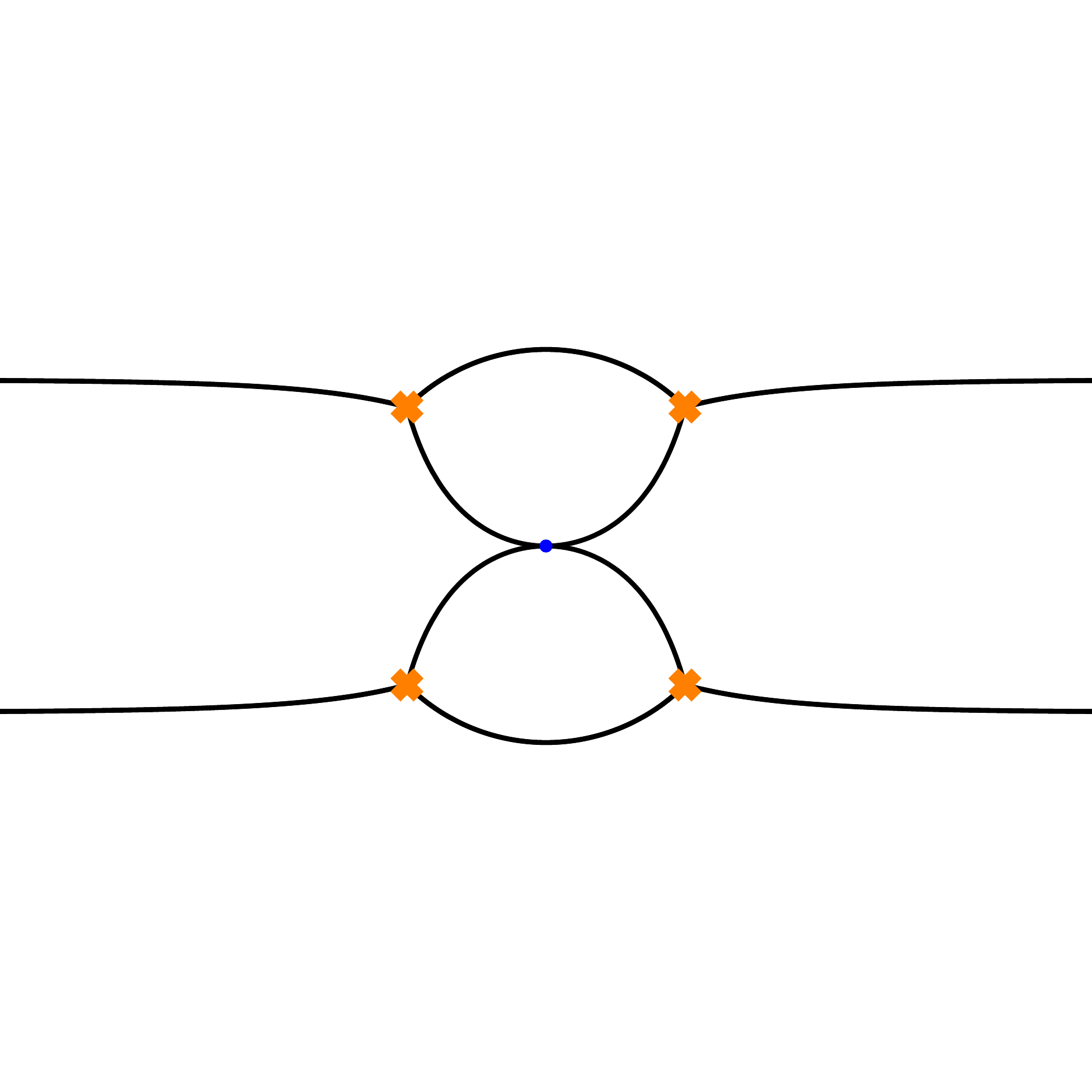}}~~~~
\resizebox{45mm}{!}{\includegraphics{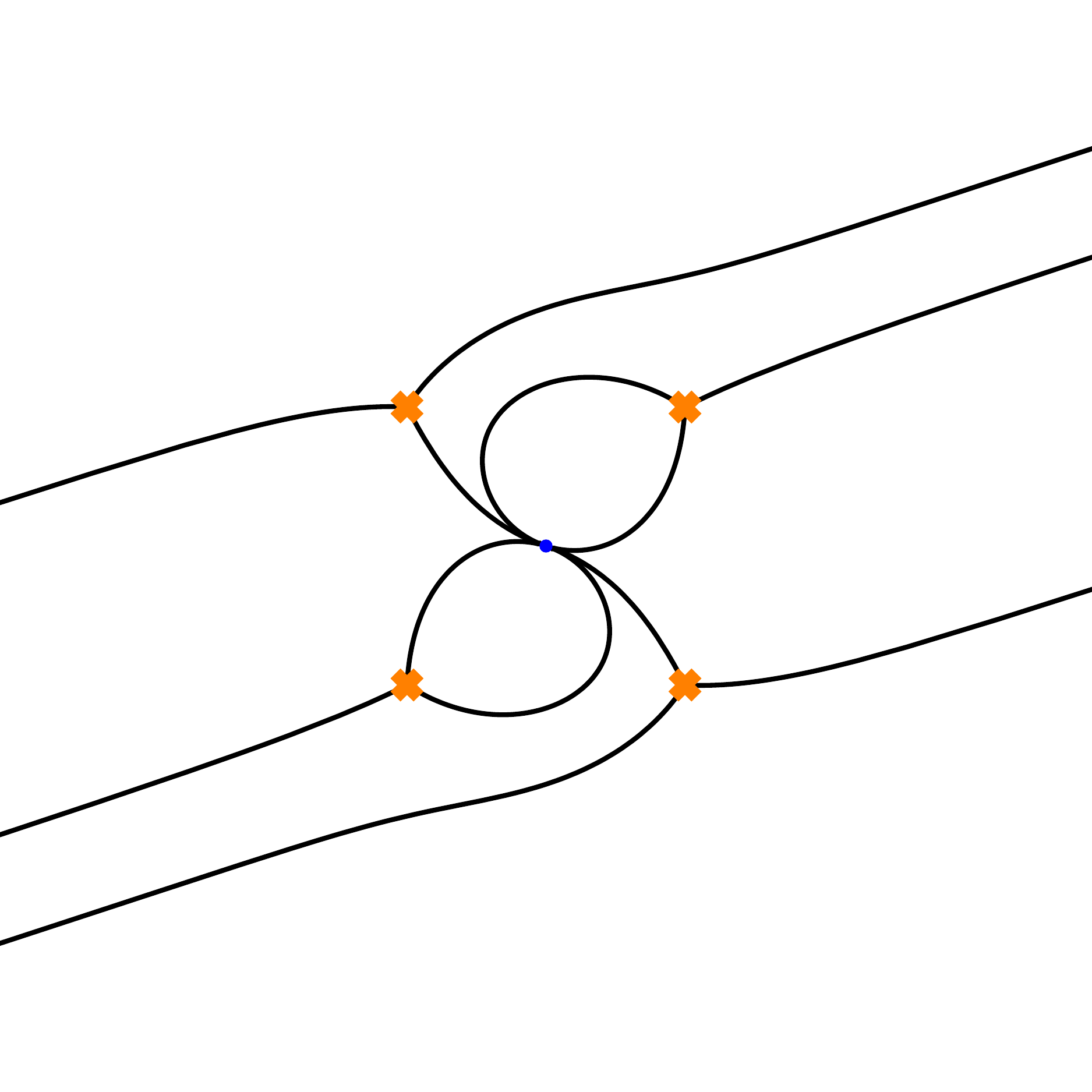}}~~~~~~~~~~~~~~
\end{center}
  \caption{The WKB curves (black lines) for $p(z)=\frac{1}{z^4}+1$. Here the spectral parameter is fixed to be $-\pi/10$, $0$ and $\pi/10$ (from left to right). The yellow crosses denote the zeros of $p(z)$. ``pop'' occurs at $\theta=0$, where the topology of the Stokes graph changes. See sec. 3.3 and appendix B in \cite{Iwaki2014} for related discussion.} 
\label{fig:mn=22}
\end{figure}

\begin{figure}[htb]
\begin{center}
\resizebox{45mm}{!}
{\includegraphics{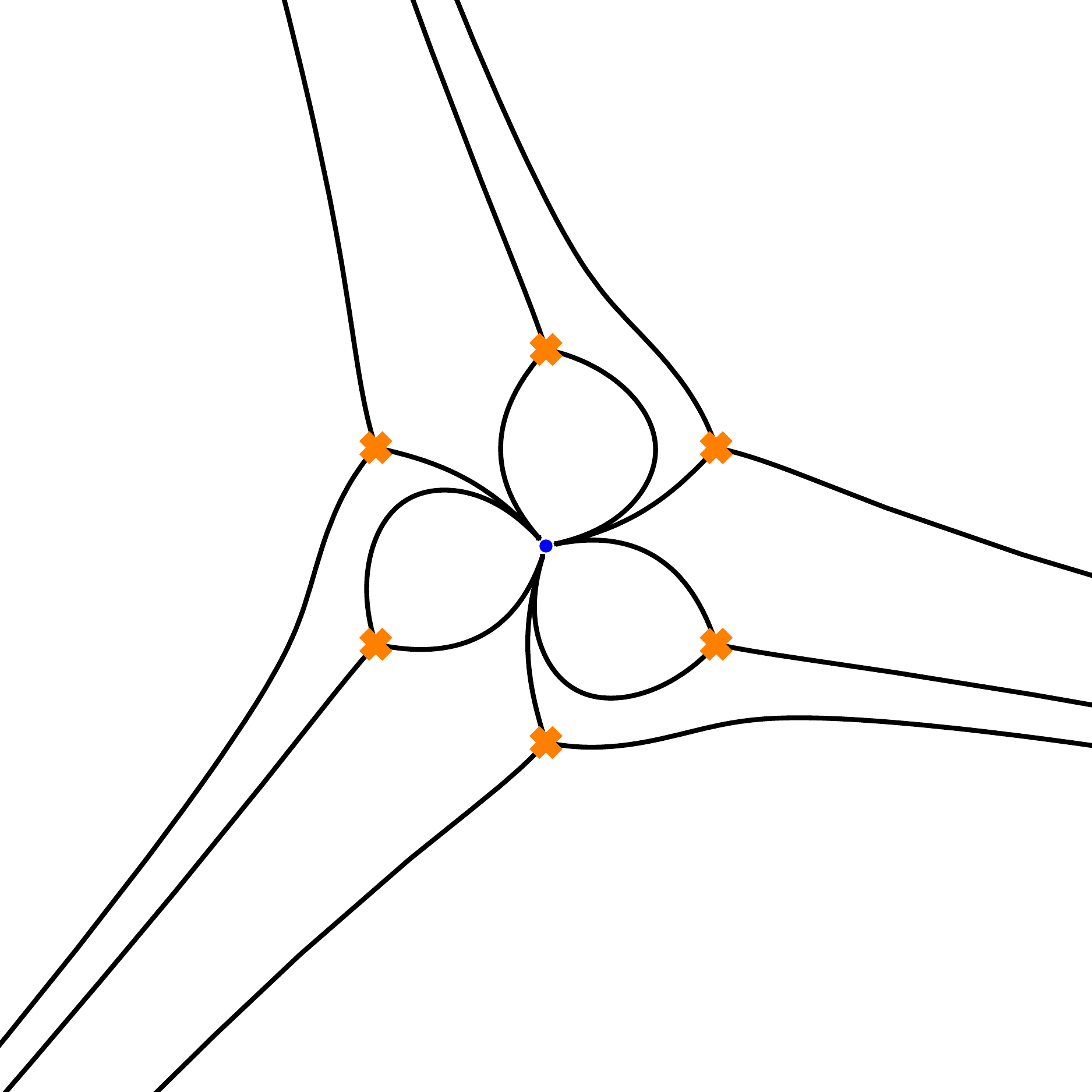}}~~~~
\resizebox{45mm}{!}{\includegraphics{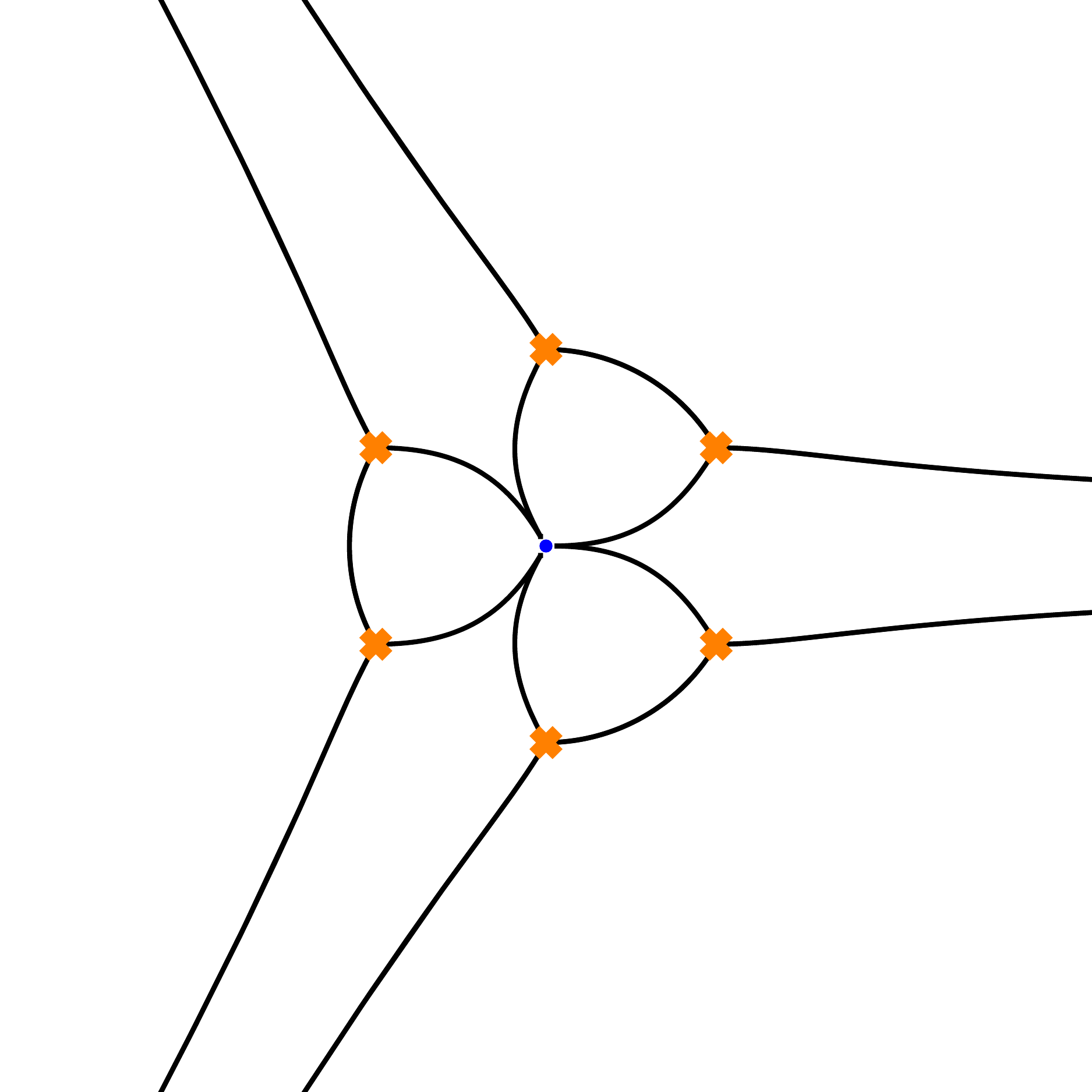}}~~~~
\resizebox{45mm}{!}{\includegraphics{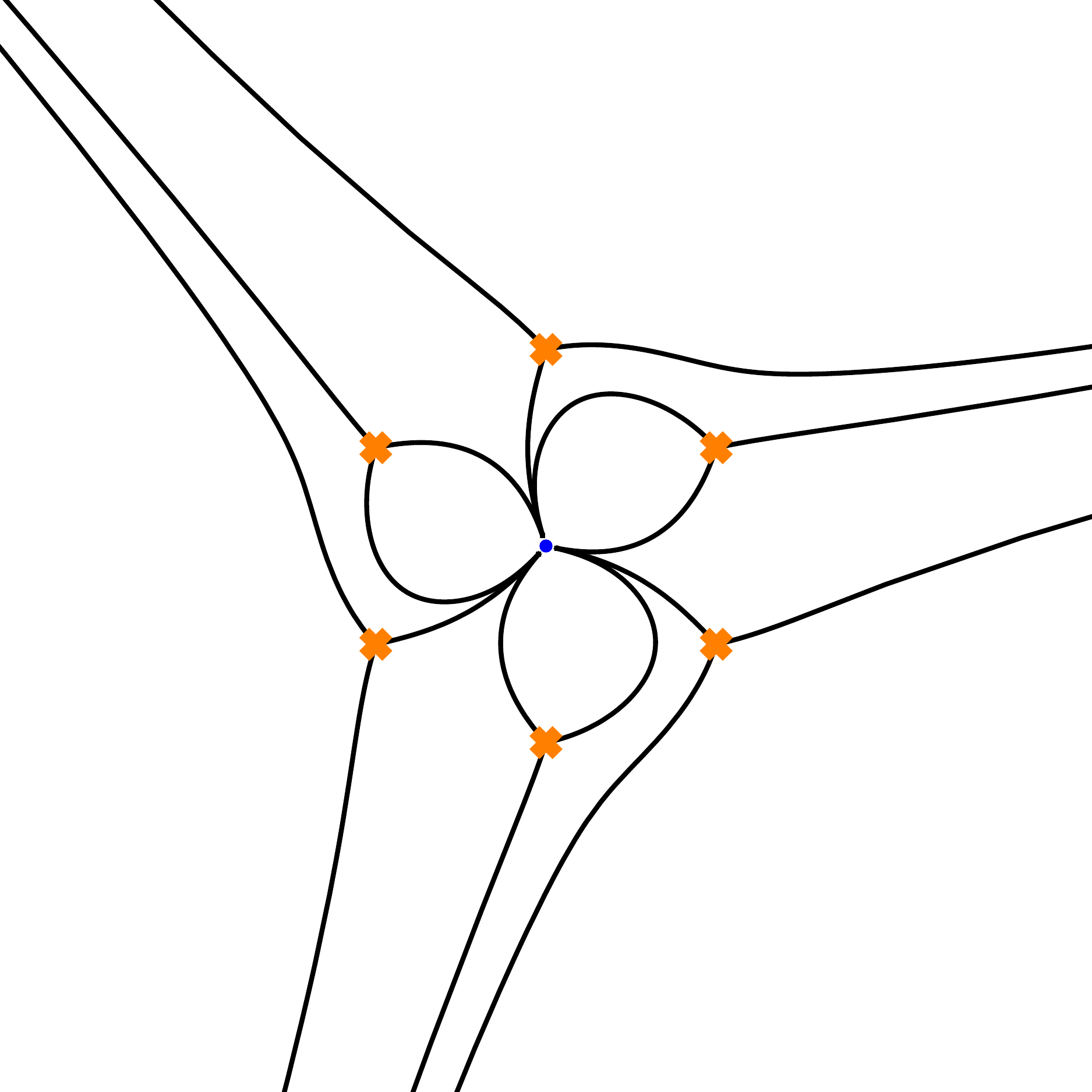}}~~~~~~~~~~~~~~
\end{center}
  \caption{The WKB curves (black lines) for $p(z)=\frac{1}{z^5}+z$. Here the spectral parameter is fixed to be $-\pi/10$, $0$ and $\pi/10$ (from left to right). The yellow crosses denote the zeros of $p(z)$.} 
\label{fig:mn=33}
\end{figure}

 Let us consider the $\zeta\to 0$ case for instance. It is convenient to diagonalize $\Phi_z\to {\rm diag}(\sqrt{p}, -\sqrt{p})$, such that the solution of linear problem behaves as $\exp(\pm \frac{1}{\zeta}\int^z_{z_\ast}\sqrt{p(z^\prime)}dz^\prime)$. It is thus natural to consider the problem on the following Riemann surface:
\be
\label{eq:Rie-sur}
y^2=p(z).
\ee
To make sure the precision of the WKB approximation, we follow the solutions along the path of WKB curve:
\be
{\rm Im}(\frac{1}{\zeta}\sqrt{p(z)}\frac{dz}{dt})=0,	
\ee	
which is parametrized by $t$. At a generic point on the complex $z$-plane, WKB curves do not intersect. At the (simple) zeros of $p(z)$, three WKB curves radiate, which separate the plane into three regions. In Fig.\ref{fig:mn=22} and Fig.\ref{fig:mn=33}, we plot the WKB curves for the Riemann surface $y^2=p(z)$ 
with $(m, n)=(2,2)$ and $(m, n)=(3,3)$ respectively. 

\begin{figure}[htb]
\begin{center}
\resizebox{75mm}{!}{\includegraphics{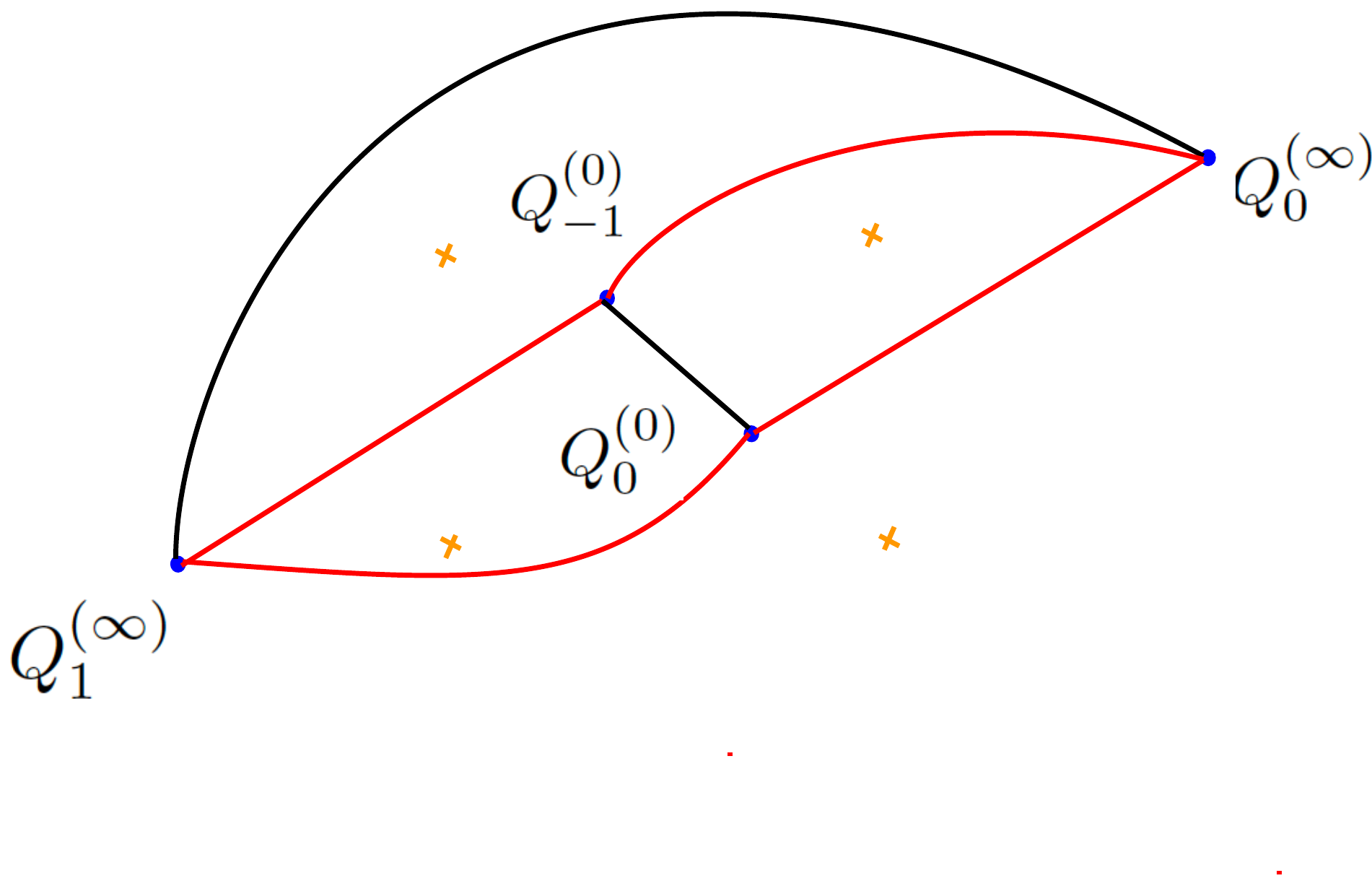}}~~~~
\resizebox{65mm}{!}{\includegraphics{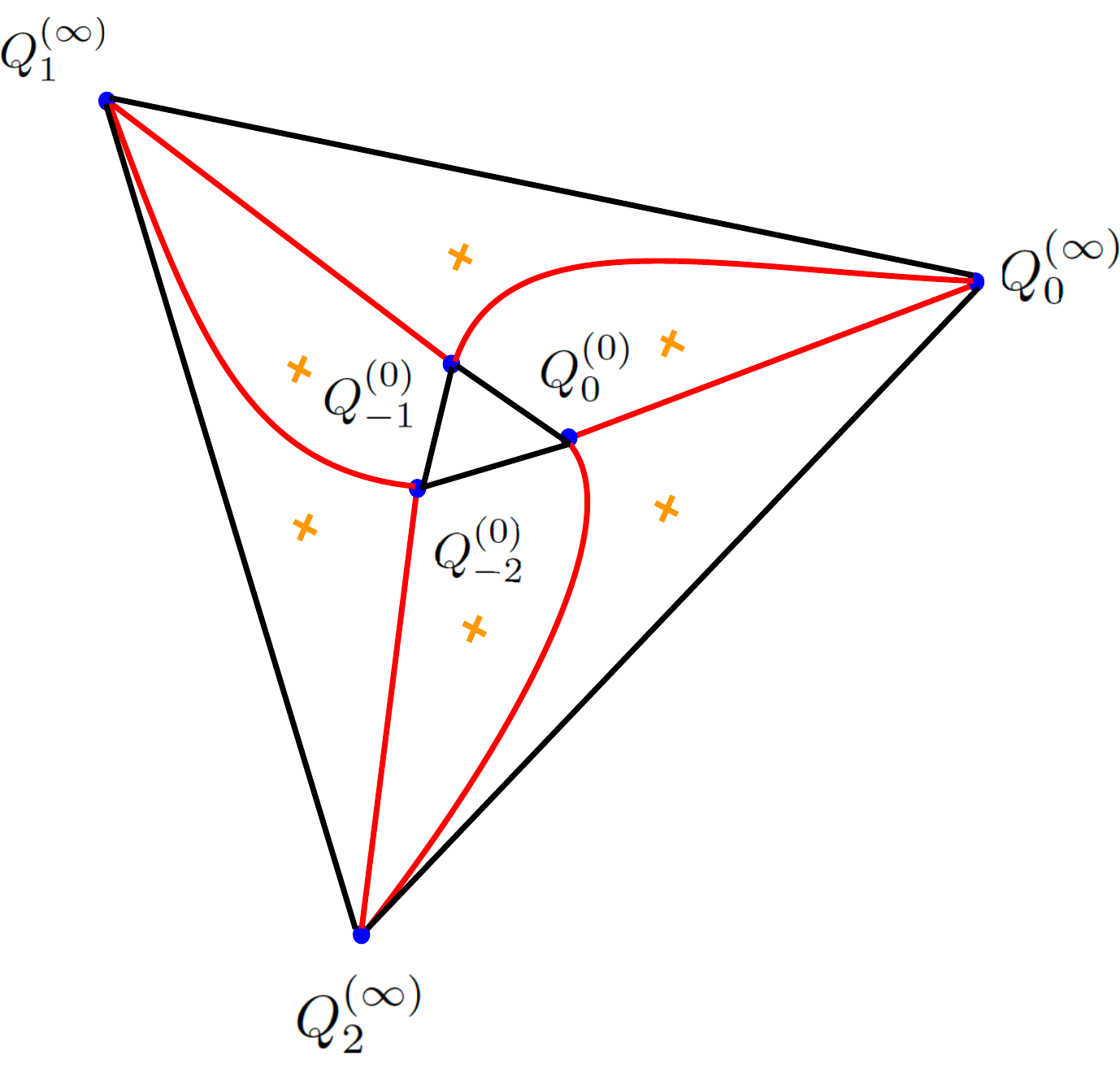}}~~~~~~~~~~~~~~
\end{center}
  \caption{The WKB triangulation for $p(z)=\frac{1}{z^4}+1$ and $p(z)=\frac{1}{z^5}+z$ with $0<{\rm Im}(\theta)<\pi/2$.} 
\label{fig:wkbtri-mn}
\end{figure}

In our case, $z=0$ and $z=\infty$ are order $m+2$ and $n+2$ irregular singular points respectively, say $Q^{(i)}$ with $(i)=(0),(\infty)$. The WKB curves emerging from $Q^{(i)}$ will divide the surface into $n^{(i)}$ sectors, where $n^{(\infty)}=n$ and $n^{(0)}=m$. It is thus convenient to regard this irregular singular point $Q^{(i)}$ as $n^{(i)}$ marked singular points $Q_{k}^{(i)}$, $k=1,\cdots,n^{(i)}$. We will locate $Q^{(i)}_k$ on the direction where the small solution $s_k^{(i)}$ decays the fastest, such that the small solution is uniquely defined around each marked point\footnote{More details can be found in section 8 of \cite{GaiottoMooreNeitzke:2009}. }. The complex plane will be divided by these WKB curves into cells.  In each cell, several homotopically equivalent curves sweep. Choosing a representative curve from each family, we obtain the WKB triangulation $T_{\rm WKB}$, which means a triangulation by the WKB curves with all vertices $Q$ at the regular singularities or the marked points of irregular points, and at least one edge $E$ ends on each vertex. Two triangles, which bound edge $E$, make up a quadrilateral ${\cal Q}_E$, where the small solution to be single-valued and smooth defined up to rescaling. In Fig.\ref{fig:wkbtri-mn}, we show the WKB triangulation $T_{\rm WKB}$ for the Riemann surface $y^2=p(z)$ for $(m, n)=(2,2)$ and $(m, n)=(3,3)$ with $0<{\rm Im}(\theta)<\pi/2$.

In Fig.\ref{fig:wkbtri-mn}, the black lines, connecting the marked points of the same irregular point, are called boundary edges, whose Fock-Goncharov coordinates are set to be zero \cite{GaiottoMooreNeitzke:2009}. The nontrivial Fock-Goncharov coordinates are the ones associated with the red lines in Fig.\ref{fig:wkbtri-mn}.

\subsection{Fock-Goncharov coordinates and functional relation}
\subsubsection{$m=n$ case}
Let us consider the case of $m=n$ at first. We consider $p(z)=c(z^{-3}+z^{-1})$ as a typical case and choose $0<\phi<\pi/2$, so the WKB triangulation is given by Fig.\ref{fig:wkbtri-mn} for $n=2,3$. In general there are $2n$ non-trivial edges, $E(Q_{-k}^{(0)}, Q_k^{(\infty)})$ and $E(Q_{-k-1}^{(0)}, Q_k^{(\infty)})$ with $k=0,1,...,n-1$. We introduce two types of Fock-Goncharov coordinates
\be
\ba
{\cal X}_{-k,k}&:={\cal X}_{E(Q_{-k}^{(0)},Q_{k}^{(\infty)})}=-\frac{(s_{k-1}^{(\infty)}\wedge s_{k}^{(\infty)})(s_{-k-1}^{(0)}\wedge s_{-k}^{(0)})}{(s_{-k}^{(0)}\wedge s_{k-1}^{(\infty)})(s_{k}^{(\infty)}\wedge s_{-k-1}^{(0)})}=\frac{1}{(s_{-k}^{(0)}\wedge s_{k-1}^{(\infty)})(s_{-k-1}^{(0)}\wedge s_{k}^{(\infty)})},\\
{\cal X}_{-k-1,k}&:={\cal X}_{E(Q_{-k-1}^{(0)},Q_{k}^{(\infty)})}=-\frac{(s_{-k}^{(0)}\wedge s_{k}^{(\infty)})(s_{k+1}^{(\infty)}\wedge s_{-k-1}^{(0)})}{(s_{-k-1}^{(0)}\wedge s_{-k}^{(0)})(s_{k}^{(\infty)}\wedge s_{k+1}^{(\infty)})}=(s_{-k}^{(0)}\wedge s_{k}^{(\infty)})(s_{-k-1}^{(0)}\wedge s_{k+1}^{(\infty)}),
\ea
\ee
which are associated with the cross ratios (\ref{eq:cross}):
\be
{\cal X}_{-k,k}(\zeta=1,i)^{-1}=-\chi_{-k,k-1,-k-1,k}^{(0,\infty,0,\infty)\pm},\quad {\cal X}_{-k-1,k}(\zeta=1,i)=-\chi_{-k,k,-k-1,k+1}^{(0,\infty,0,\infty)\pm}.
\ee
Different coordinates are related by $\mathcal{X}_{-k,k}=\mathcal{X}_{-k-n,k+n}$ and $\mathcal{X}_{-k-1,k}=\mathcal{X}_{-k-1-n,k+n}$ because $s^{(0)}_{-k}=M s^{(0)}_{-n-k}$ and $s^{(\infty)}_{k}=M s^{(\infty)}_{n+k}$, where $M$ is the monodromy operator around $z=0$.

For any edge $E(P,Q)$, it is convenient to introduce a function
\be\label{eq:APQ}
A_{PQ}=-\frac{(s_{Q}\wedge\tilde{s}_{P})(s_{P}\wedge\tilde{s}_{Q})}{(s_{P}\wedge\tilde{s}_{P})(s_{Q}\wedge\tilde{s}_{Q})},
\ee
where $\tilde{\cdot}$ denote the $\mathbb{Z}_2$ shift of the solutions. For the small solution, $\tilde{s}_k^{(i)}=s_{k+1}^{(i)}$. By using the Pl\"ucker relation (Schouten relation) for any vectors $v_1,\cdots,v_4$:
\be
(v_{1}\wedge v_{2})(v_{3}\wedge v_{4})+(v_{3}\wedge v_{1})(v_{2}\wedge v_{4})+(v_{2}\wedge v_{3})(v_{1}\wedge v_{4})=0,
\ee
one finds
\be
1+A_{PQ}=\frac{(s_{P}\wedge s_{Q})(\tilde{s}_{P}\wedge\tilde{s}_{Q})}{(s_{P}\wedge\tilde{s}_{P})(s_{Q}\wedge\tilde{s}_{Q})}.
\ee
Since $1+A_{Q^{(i)}_kQ^{(i)}_{k+1}}=1$, only $A_{k,j}:=A_{Q_{k}^{(0)}Q_{j}^{(\infty)}}$ is non-trivial. It is then easy to find
\be
\label{eq:m=n-rel}
\ba
{\cal X}_{-k,k}\tilde{{\cal X}}_{-k,k}=\frac{1}{\big(1+A_{-k,k-1}\big)\big(1+A_{-k-1,k}\big)},\quad{\cal X}_{-k-1,k}\tilde{{\cal X}}_{-k-1,k}=\big(1+A_{-k,k}\big)\big(1+A_{-k-1,k+1}\big).
\ea
\ee
The right hand side can be expressed in terms of coordinates by noting
\be
\label{eq:A-X-rel}
\ba
	A_{-k,k}=&\mathcal{X}_{-k,k}(1+\mathcal{X}_{-k,k-1})(1+\mathcal{X}_{-k-1,k}),\\
	A_{-k,k-1} =&\mathcal{X}_{-k,k-1},
\ea
\ee
which are easily derived by using the Pl\"ucker relation. Together with the conditions $\mathcal{X}_{-i,i}=\mathcal{X}_{-i-n,i+n}$ and $\mathcal{X}_{-i-1,i}=\mathcal{X}_{-i-1-n,i+n}$, we obtain a closed system with $2n$ coordinates. In the following, we show the case $(m,n)=(1,1), (2,2)$ for instance:
\paragraph{Example: $(m,n)=(1,1)$}
\be
\label{eq:mn=11-rel}
\ba
{\cal X}_{0,0}\tilde{{\cal X}}_{0,0}&=\frac{1}{\big(1+\mathcal{X}_{-1,0}\big)^{2}},\\
{\cal X}_{-1,0}\tilde{{\cal X}}_{-1,0}&=\big(1+\mathcal{X}_{0,0}(1+\mathcal{X}_{-1,0})^{2}\big)^{2}.
\ea
\ee

\paragraph{Example: $(m,n)=(2,2)$}
\be
\ba
{\cal X}_{0,0}\tilde{{\cal X}}_{0,0}&=\frac{1}{\big(1+\mathcal{X}_{-2,1}\big)\big(1+\mathcal{X}_{-1,0}\big)},\\
{\cal X}_{-1,0}\tilde{{\cal X}}_{-1,0}&=\big(1+\mathcal{X}_{0,0}(1+\mathcal{X}_{-2,1})(1+\mathcal{X}_{-1,0})\big)\big(1+\mathcal{X}_{-1,1}(1+\mathcal{X}_{-1,0})(1+\mathcal{X}_{-2,1})\big),\\
{\cal X}_{-1,1}\tilde{{\cal X}}_{-1,1}&=\frac{1}{\big(1+\mathcal{X}_{-1,0}\big)\big(1+\mathcal{X}_{-2,1}\big)},\\
{\cal X}_{-2,1}\tilde{{\cal X}}_{-2,1}&=\big(1+\mathcal{X}_{-1,1}(1+\mathcal{X}_{-1,0})(1+\mathcal{X}_{-2,1})\big)\big(1+\mathcal{X}_{0,0}(1+\mathcal{X}_{-2,1})(1+\mathcal{X}_{-1,0})\big).
\ea
\ee

\subsubsection{$n>m$ case}
\begin{figure}[tbh]
\begin{center}
\resizebox{120mm}{!}{\includegraphics{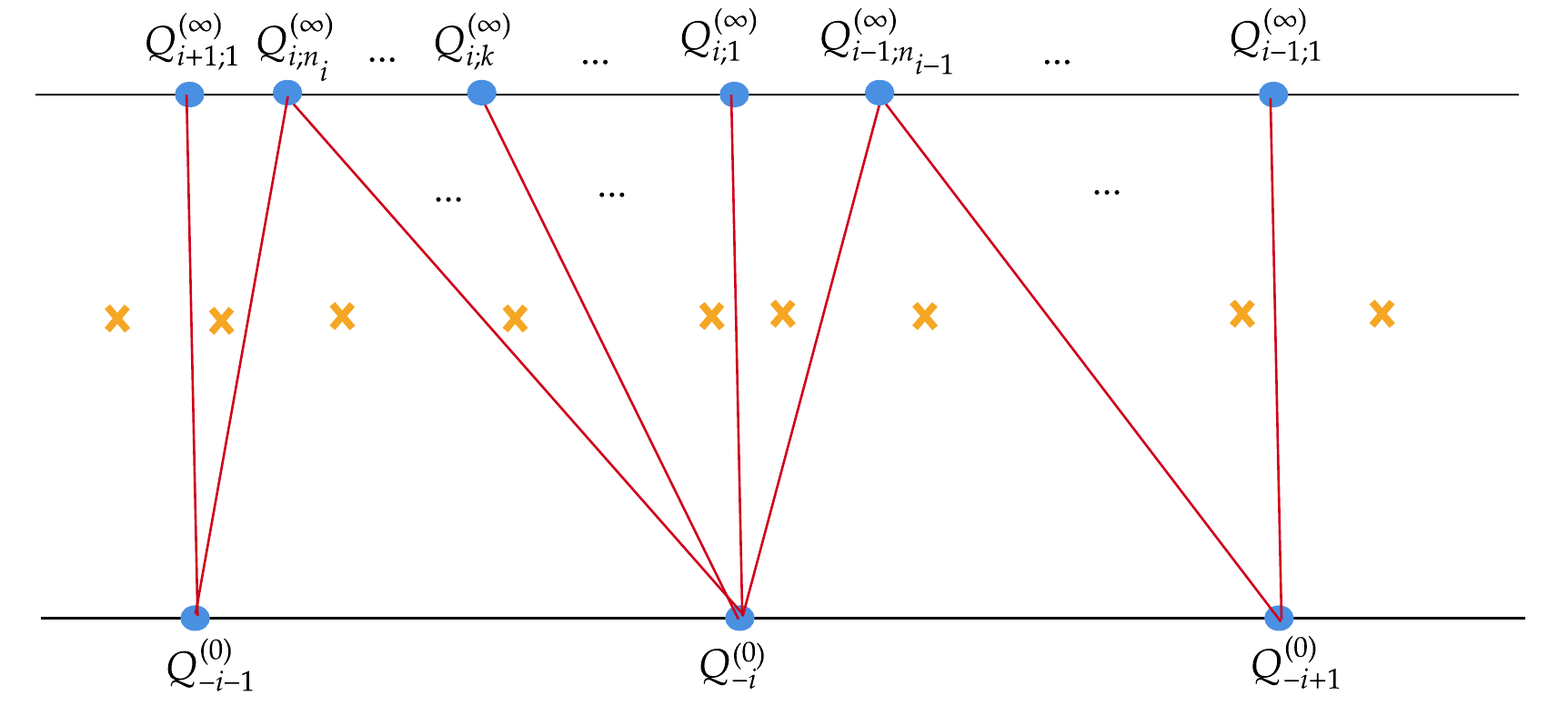}}~~~~
\end{center}
  \caption{The WKB triangulation for $n>m$.} 
\label{fig:nlm}
\end{figure}

We then consider the case $n>m$. It is convenient to partition the $n$ Stokes sectors into $m$ groups corresponding to the WKB triangulation as shown in Fig. \ref{fig:nlm}. The $i$th group contains $n_i$ Stokes sectors $S_{i;k}^{(\infty)}$, $k=1,...,n_i$  which connect to $\mathcal S_{-i}^{(0)}$ through WKB curves. In addition, $\mathcal S_{i;n_i}^{(\infty)}$ and $\mathcal S_{-i-1}^{(0)}$ are also connected so we define $\mathcal S_{i+1;0}^{(\infty)}=\mathcal S_{i;n_i}^{(\infty)}$. We define the Fock-Goncharov coordinates as
\be
\ba
{\cal X}_{i;k}&:={\cal X}_{E(Q_{-i}^{(0)},Q_{i;k}^{(\infty)})}=-\frac{(s_{i;k-1}^{(\infty)}\wedge s_{i;k}^{(\infty)})(s_{i;k+1}^{(\infty)}\wedge s_{-i}^{(0)})}{(s_{-i}^{(0)}\wedge s_{i;k-1}^{(\infty)})(s_{i;k}^{(\infty)}\wedge s_{i;k+1}^{(\infty)})}, ~~~k=1,...,n_i-1,\\
{\cal X}_{i;n_i}&:={\cal X}_{E(Q_{-i}^{(0)},Q_{i;n_i}^{(\infty)})}=-\frac{(s_{i;n_i-1}^{(\infty)}\wedge s_{i;n_i}^{(\infty)})(s_{-i-1}^{(0)}\wedge s_{-i}^{(0)})}{(s_{-i}^{(0)}\wedge s_{i;n_i-1}^{(\infty)})(s_{i;n_i}^{(\infty)}\wedge s_{-i-1}^{(0)})},\\
{\cal X}_{i;0}&:={\cal X}_{E(Q_{-i}^{(0)},Q_{i-1;n_{i-1}}^{(\infty)})}=-\frac{(s_{-i+1}^{(0)}\wedge s_{i-1;n_{i-1}}^{(\infty)})(s_{i;1}^{(\infty)}\wedge s_{-i}^{(0)})}{(s_{-i}^{(0)}\wedge s_{-i+1}^{(0)})(s_{i-1;n_{i-1}}^{(\infty)}\wedge s_{i;1}^{(\infty)})}.\\ 
\ea
\ee
The functional relations are
\begin{equation}
\begin{aligned}
  {\cal X}_{i;k} {\cal \tilde X}_{i;k}
  &=\frac{1+A_{i;k+1}}{1+A_{i;k-1}}
  , ~~~k=1,...,n_i-1,\\
    {\cal X}_{i;n_i} {\cal \tilde X}_{i;n_i}
  &=\frac{1}{(1+A_{i;n_i-1})(1+A_{i+1;0})},\\
    {\cal X}_{i;0} {\cal \tilde X}_{i;0}
  &=(1+A_{i-1;n_{i-1}})(1+A_{i;1}),
\end{aligned}
\end{equation}
where
\begin{equation}
\begin{aligned}
  A_{i;k} 
  &={\cal X}_{i;k}(1+{\cal X}_{i;k-1}(1+...{\cal X}_{i;1}
  (1+{\cal X}_{i;0}))), ~~~k=1,...,n_i-1
  \\
    A_{i;n_i} 
  &
  ={\cal  X}_{i;n_i}(1+{\cal X}_{i;n_i-1}(1+...{\cal X}_{i;1}
  (1+{\cal X}_{i;0})))
  (1+{\cal X}_{i+1;0}),
  \\
    A_{i;0} 
  &
  ={\cal \tilde X}_{i;0}.
\end{aligned}
\end{equation}

As a simple example, the functional relations for $(m,n)=(1,2)$ are:
\paragraph{Example: $(m,n)=(1,2)$}
\be
\ba
{\cal X}_{0;0}\tilde{{\cal X}}_{0;0}&=
\big(1+\mathcal{X}_{0;2}(1+\mathcal{X}_{0;1}(1+\mathcal{X}_{0;0}))(1+\mathcal{X}_{0;0})\big)
\big(1+\mathcal{X}_{0;1}(1+\mathcal{X}_{0;0})\big),\\
{\cal X}_{0;1}\tilde{{\cal X}}_{0;1}&=\frac{\big(1+\mathcal{X}_{0;2}(1+\mathcal{X}_{0;1}(1+\mathcal{X}_{0;0}))(1+\mathcal{X}_{0;0})\big)}{\big(1+\mathcal{X}_{0;0}\big)},\\
{\cal X}_{0;2}\tilde{{\cal X}}_{0;2}&=\frac{1}{\big(1+\mathcal{X}_{0;1}(1+\mathcal{X}_{0;0})\big)
\big(1+\mathcal{X}_{0;0}\big)}.
\ea
\ee

\subsection{TBA-like equations}
\begin{figure}[tbh]
\begin{center}
\resizebox{120mm}{!}{\includegraphics{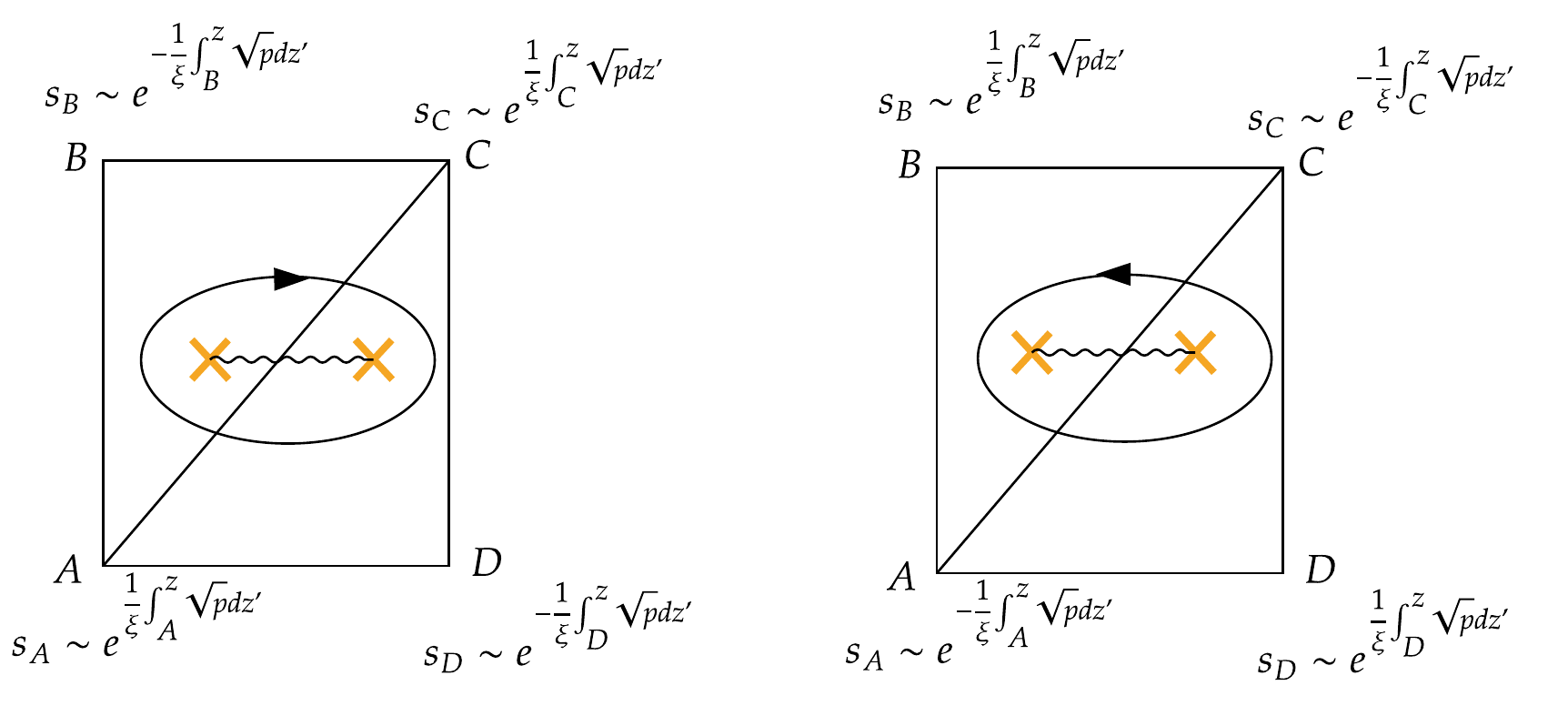}}~~~~
\end{center}
  \caption{The cycle $\gamma_{E(A,C)}$  associated with the Fock-Goncharov coordinates $\mathcal{X}_{E(A,C)}$.  Two cases are distinguished according to the signs in front of the exponential parts of the small solutions.} 
\label{fig:gammaE}
\end{figure}

The standard WKB approximation shows 
\be
\log{\cal X}^{(0)}_E=\frac{1}{\zeta}Z_E+\zeta \bar{Z}_E=\frac{1}{\zeta}\oint_{\gamma_E}\sqrt{p(z)} dz+\zeta\oint_{\gamma_E}\sqrt{\bar{p}(\bar{z})} d\bar{z},\label{asymptotics}
\ee
for $\zeta\to 0, \infty$, where $\gamma_E$ is the cycle encircling the two zeros in the quadrilateral $\mathcal Q_E$ \cite{GaiottoMooreNeitzke:2009}. The definition of $\gamma_E$ is shown in Fig.\ref{fig:gammaE}.
Let us denote the right hand side of the the functional relations by ${\cal F}_{E_i}$, i.e.
\be
\label{fun-rel}
{\cal X}_{i}\tilde{{\cal X}}_{i}=:{\mathcal F_i},
\ee 
where $i$ is the label of the edge $E_i$. This relation can be inverted into TBA-like equations by using the Fourier transformation
\be
\ba\label{Ieq-x}
\log{\cal X}_{i}(\theta+i\phi)=\log{\cal X}_{i}^{(0)}(\theta+i\phi  )+\int_{\mathbb{R}}\frac{d\theta^{\prime}}{2\pi}\frac{\log{\cal F}_i(\theta^{\prime}-\frac{\pi i}{2}+i\phi)}{\cosh(\theta-\theta^{\prime})},
\ea
\ee
or equivalently
\be
\ba\label{Ieq-xA}
\log{\cal X}_{i}(\theta+i\phi-\frac{\pi i}{2})=\log{\cal X}_{i}^{(0)}(\theta+i\phi -\frac{\pi i}{2})-\int_{\mathbb{R}}\frac{d\theta^{\prime}}{2\pi i}\frac{\log{\cal F}_i(\theta^{\prime}-\frac{\pi i}{2}+i\phi)}{\sinh(\theta-\theta^{\prime}+i\epsilon)},
\ea
\ee
where ${\cal X}_{i}^{(0)}(\theta)$ denotes the leading order of ${\cal X}_{i}(\theta)$ which can be read from \eqref{asymptotics}. $\epsilon$ is a small positive number. Note that these formulas are valid when the leading order of coordinates, ${\cal X}_i^{(0)}(\theta+i\phi)=\exp\big(e^{-\theta-i\phi}Z_{E_i}+e^{\theta+i\phi}\bar{Z}_{E_i}\big)$, are convergent for $\theta\to \pm \infty$, which is the main scope in this paper.

\subsection{Area from the TBA-like equation}

\begin{figure}[h]
\begin{center}
\resizebox{120mm}{!}{\includegraphics{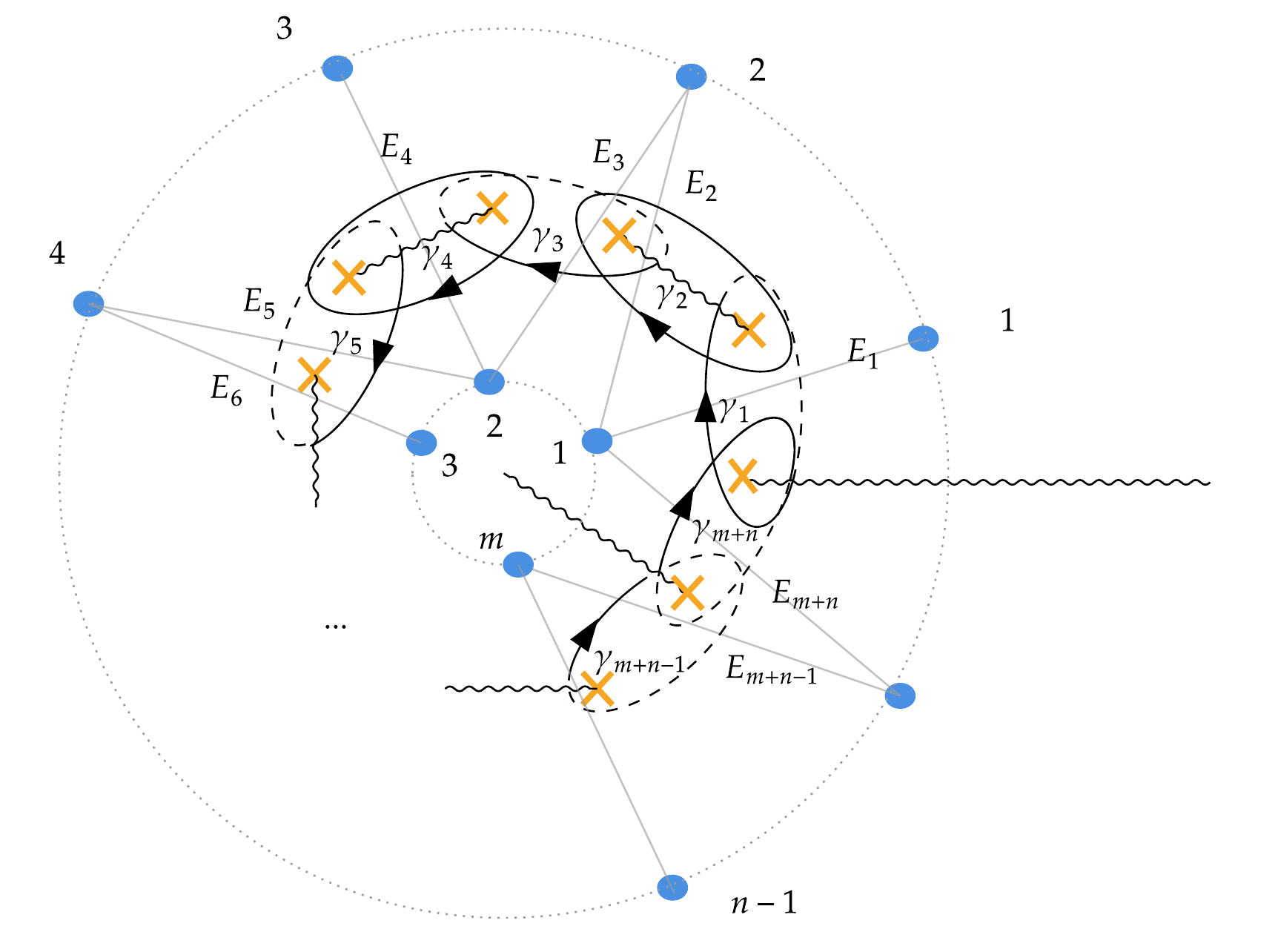}}~~~~
\end{center}
  \caption{ The cycles for Riemann bilinear identity for odd $n$ and $m$.}
\label{fig:rienm}
\end{figure}
The area $A_{\rm reg}$ can be reduced  to one-dimensional integrals over cycles by applying the Riemann bilinear identity:
\begin{equation}\label{Areg}
	A_{\rm reg}=\int dz^2 \sqrt{p} u
	=
	\frac{i}{4}\int_\Sigma \sqrt{p} dz
	\wedge \eta
	=- \frac{i}{4}\sum_{i=1}^{{n+m}}\oint_{C_i}
	\sqrt{p}  \eta
	- \frac{i}{4}\oint_{\gamma_a}\sqrt{p} dz I^{ab}\oint_{\gamma_b} \eta.
\end{equation}
where $C_i$ is a small contour encircling the zero point $z_i$ and  $\{\gamma_a\}$ is a complete basis of cycles. The matrix $I^{ab}$ is the inverse of intersection matrix of the cycles.
We denote by $\partial_a$ the tangent vector of the  cycle $\gamma_a$.
For each pair of intersecting cycles $\gamma_a$ and $\gamma_b$,
we have $I_{ab}=1(-1)$ if
$\det[\{\partial_a,\partial_b \}] > 0(< 0)$ at the intersecting point.

Using the explicit expression
\be
 \eta=2\sqrt{\bar{p}}\big(\cosh(2\hat{\alpha})-1\big)d\bar{z}+\frac{1}{\sqrt{p}}(\partial\hat{\alpha})^{2} dz,
 \ee
the contribution from a $C_i$ can be computed as
\begin{equation}
	- \frac{i}{4}\int_{C_i} \sqrt{p} \eta   = \frac{\pi}{24}.
\end{equation}
When both $n$ and $m$ are odd integers, we choose the basis of cycles $\{\gamma_a\}$ as shown in fig. \ref{fig:rienm}. Depending on the signs in front of the exponential parts of the small solutions of the Stokes sectors, some of the $\gamma_a$ defined here are in the opposite direction of  the $\gamma_E$ used in the asymptotics (\ref{asymptotics}) of the $\mathcal{X} $ coordinates.
In our convention, the signs are chosen to be:
\be
s_{k}^{(\infty,0)}\sim  \exp \left(  (-)^k\frac{1}{\zeta}\int_{Q_k^{(\infty,0)}}^{z}dz'\sqrt{p}+...  \right),
\ee
which is consistent with the Stokes sector defined in (\ref{eq:ssectorinf}) and (\ref{eq:ssector0}). For even (resp. odd) $k$,  $\gamma_{E(Q_k^{(0)},*)}$ is in the same (resp. opposite) direction with respect to the associated $\gamma_a$ defined in fig. \ref{fig:rienm}.

The nonzero elements of the intersection matrix $I_{ab}$ are
\begin{equation}
	I_{n+m,n+m-1}=-I_{n+m-1,n+m}=I_{n+m,1}=-I_{1,n+m}=I_{a,a+1}=-I_{a+1,a}=1,
\end{equation}
where $a=1,...,n+m-2$.
The integrals of $\eta$ over the $\gamma$-cycles can be written as:
\begin{equation}
\oint_{\gamma_a} \eta= \sum_{b=1}^{n+m} I_{a b}\int_{\beta_b} \eta,
\end{equation}
where  $\beta_a$ is the path from $0$ to $\infty$ that intersects $\gamma_a$.
The area can be simplified as
\begin{equation}\label{eq:areg-rie}
A_{\mathrm{reg}}=\frac{\pi}{24} (n+m)-\frac{i}{4}	\sum_{a=1}^{n+m} \int_{\beta_a} \eta \oint_{\gamma_{a}}\sqrt{p} dz .
\end{equation}
One can further show that (\ref{eq:areg-rie}) holds for general values of $n$ and $m$.

Let us denote the fastest decay solution of the linear problem at puncture $Q_i$ by $s_i$, say the small solution. As shown in \cite{Kazama:2011cp}, the information of $\lambda$ and $ud\bar{z}+v dz$ on the path $Q_i\to Q_j$ can be extracted from the WKB approximation of $\log s_i\wedge s_j$.
When there is a WKB curve connecting $Q_i$ and $Q_j$, the signs in front of the WKB expansion of the exponential parts of $s_i$ and $s_j$ must be opposite \footnote{$Q_i$ and $Q_j$ can be the mark points of different irregular singular points.}. To be concrete, if
\be
\ba
s_{i}\sim& \exp \left(  \frac{1}{\zeta}\int_{Q_i}^{z}dz'\sqrt{p}+\zeta\Big(\frac{1}{2}\int_{Q_i}^{z}dz'\big(\frac{(\partial\hat{\alpha})^{2}}{\sqrt{p}}-\partial(\frac{\partial\hat{\alpha}}{\sqrt{p}})\big)+\int_{Q_i}^{\bar z}d\bar{z}'\sqrt{\bar{p}}e^{-2\hat{\alpha}}\Big)   \right),\\
s_{j}\sim& \exp \left(-  \frac{1}{\zeta}\int_{Q_j}^{z}dz'\sqrt{p}-\zeta\Big(\frac{1}{2}\int_{Q_j}^{z}dz'\big(\frac{(\partial\hat{\alpha})^{2}}{\sqrt{p}}-\partial(\frac{\partial\hat{\alpha}}{\sqrt{p}})\big)-\int_{Q_j}^{\bar z}d\bar{z}'\sqrt{\bar{p}}e^{-2\hat{\alpha}}\Big)   \right),
\ea
\ee
the WKB expansion of $\log s_{i}\wedge s_{j}$ is
\be
\ba\label{wskss}
\log s_{i}\wedge s_{j}&\sim\frac{1}{\zeta}\int_{Q_i}^{Q_j}dz\sqrt{p}+\zeta\Big(\frac{1}{2}\int_{Q_i}^{Q_j}dz\big(\frac{(\partial\hat{\alpha})^{2}}{\sqrt{p}}-\partial(\frac{\partial\hat{\alpha}}{\sqrt{p}})\big)+\int_{Q_i}^{Q_j}d\bar{z}\sqrt{\bar{p}}e^{-2\hat{\alpha}}\Big)+\cdots\\
&\sim\frac{1}{\zeta}\int_{Q_i}^{Q_j}\sqrt{p} dz+\zeta\int_{Q_i}^{Q_j}d\bar{z}\sqrt{\bar{p}}+\frac{\zeta}{2}\int_{Q_i}^{Q_j}(ud\bar{z}+vdz)+\cdots.
\ea
\ee
We thus can extract the information of $\lambda$ and $ud\bar{z}+v dz$ around certain cycles from the WKB approximation of $s_{i}\wedge s_{j}$.

 We now show that $\eta_i$ can be extracted from $A_i$.
Using the normalization $s_Q\wedge \tilde s_Q=1=s_P\wedge \tilde s_P$, we have
\begin{equation}
\label{eq:sA-rel}
\log \big(s_{P}\wedge  s_Q\big)^+
+
\log  \big(s_{P}\wedge  s_Q\big) ^-
=\log(1+A_{PQ}^-),
\end{equation}
which is the logarithm of \eqref{eq:APQ}. Performing the Fourier transforms at $\theta+i\phi$, we obtain
\begin{equation}
\log \big(s_{P}\wedge  s_Q\big)
(\theta+i \phi)
=\ell_{PQ}(\theta+i\phi )+\int_{\mathbb R} \frac{d\theta'}{2\pi}\frac{\log\big(1+A_{PQ}^-(\theta'+i \phi)\big)}{\cosh(\theta-\theta')},
\end{equation}
where $\ell_{PQ}(\theta)$ denotes the  leading order of $\log(s_{P}\wedge s_{Q})(\theta)$ at large $\theta$
\begin{equation}
\begin{split}
\ell_{PQ}=&\lim_{
\substack{(w,\bar w)\rightarrow Q\\ (z,\bar z)\rightarrow P}}
\bigg(\frac{1}{2}e^{-\theta-i\phi}\int_{z}^{w}\sqrt{p(z')}dz'+\frac{1}{2}e^{\theta+i\phi}\int_{\bar z}^{\bar w}\sqrt{\bar{p}(\bar{z}')}d\bar{z}'+... \bigg).
\end{split}
\end{equation}
Expanding around $\theta \rightarrow -\infty$ and comparing with (\ref{wskss}) with $\zeta=e^{\theta+i\phi}$ we find
\begin{equation}\label{eta}
    \int_{E_i} \eta=
   2\int_{\mathbb R} \frac{d\theta'}{\pi}e^{-\theta'-i\phi}{\log\big(1+A_i^-(\theta'+i \phi)\big)}.
\end{equation}
Taking into account the direction of the WKB lines, $A_{\rm reg}$ thus can be expressed as
\be
A_{\mathrm{reg}}=\frac{\pi}{24}(m+n)-\frac{i}{2}\sum_{i=1}^{m+n}\int_{\mathbb{R}}\frac{d\theta}{\pi}Z_{E_i}e^{-\theta-i\phi}\log\big(1+A_{i}^{-}(\theta+i\phi)\big).
\ee
So far, we focused on the WKB expansion around $\zeta=0$. The WKB expansion around $\zeta=\infty$ can be done in a similar way. We average the results from $\zeta=0$ and $\zeta=\infty$, and obtain
\be
A_{\mathrm{reg}}=\frac{\pi}{24}(m+n)-\frac{i}{2}\sum_{i=1}^{m+n}\int_{\mathbb{R}}\frac{d\theta}{2\pi}(Z_{E_i}e^{-\theta-i\phi}-\bar{Z}_{E_i}e^{\theta+i\phi})\log\big(1+A_{i}^{-}(\theta+i\phi)\big).
\ee
Let $\phi_i$ denotes the phase of $Z_i$ and we thus find
\be
A_{\mathrm{reg}}=\frac{\pi}{24}(m+n)+\frac{i}{2}\sum_{i=1}^{m+n}\int_{\mathbb{R}}\frac{d\theta}{\pi}|Z_{E_i}|\sinh(\theta+i\phi-i\phi_i)\log\big(1+A_{i}(\theta+i\phi-i \frac{\pi}{2})\big).
\ee
 If the WKB triangulation of interest exists at $\mathrm{Im}\,\theta=\phi_i-\frac{\pi}{2}$ for each $i$, we can choose $\phi=\phi_i-\frac{\pi}{2}$ in each integral and get
\be
\label{Areg-TBA}
A_{\mathrm{reg}}=\frac{\pi}{24}(m+n)+\sum_{i=1}^{m+n}\int_{\mathbb{R}}\frac{d\theta}{2\pi}|Z_{E_i}|\cosh\theta\log\big(1+A_{i}(\theta+i\phi_i-i \pi)\big),
\ee
where the second term has the form of free energy of the TBA-like equations (\ref{Ieq-x}). It is worth to note that the area does not depend on the value of $\zeta$ explicitly. One can introduce the Fock-Goncharov coordinates for other values of $\zeta$, which may have different Stokes graph/WKB triangulation as shown in Fig.\ref{fig:mn=22} and Fig.\ref{fig:mn=33}. This will provide the same form of the area eventually.  In appendix \ref{sec:m=n}, we will present a simplified functional relations and TBA equations for the case $m=n$, where the connection with the ${\cal N}=2$ super Yang-Mills theory is also mentioned. Two important remarks are in order here.

First, so far we have considered the function $p(z)$ as the input of the problem. It appears that the TBA-like equations and the area $A_{\rm reg}$ \eqref{Areg-TBA} do not depend explicitly on the data of the Wilson lines configuration such as the period $q$, distance $l$, and the physical cross ratios. It is useful to eliminate the central charge $Z_i$ in favor of cross ratios ${\cal X}(\zeta=1,i)$, which will depends on the physical data explicitly. Second, an arbitrary function $p(z)$  in general corresponds to two non-periodic Wilson lines. The cross ratios of the two periodic Wilson lines can be expressed in terms of ratios $(k_i/l)^\pm$. Because of momentum conservation, only $2(m+n-1)$ of such ratios are independent\footnote{This degree of freedom can also be obtained by counting the symmetries as $2(m+n)+2-3-1$. Here $2(m+n)$ is the degrees of the Wilson lines. The $2$ is the degrees of the momentum $\ell$. The $3$ is the Poincare symmetry. The $1$ is the scaling symmetry.}. But we have $(m+n)$ $\mathcal{X}$-functions (resp. $2(m+n)$ ratios). Therefore only special $p$-functions correspond to periodic Wilson lines \footnote{We are grateful to Gang Yang for pointing out this.}. We will postpone detailed  discussion of these two points to the next section.

Even though the approach presented in this section does not directly solve the physical problem, it is still important to relax the periodicity constraint momentarily and test the correctness of TBA-like equations (\ref{Ieq-xA}).

\subsection{Numeric test}
To test our TBA-like equations method,  we compare $A_{\rm reg}$ (\ref{Areg-TBA}) with the area (\ref{Areg-org}) computed by solving the generalized sinh-Gordon equation numerically. Note that the numeric test in this subsection is implemented with a given $p(z)$, which is not necessary to  correspond to a periodic Wilson lines configuration.  

It is convenient to introduce suitable function to solve for:
\be
\hat{\alpha}_{{\rm reg}}=\alpha-\frac{1}{4} \log \big((\bar z z)^{-m-2}+(\bar z z)^{n-2}\big).
\ee
It satisfies the boundary condition $\hat{\alpha}_{{\rm reg}}\to 0$ at $|z|\to 0, \infty$ and there are no  singularities at finite $|z|$. We thus can solve the generalized sinh-Gordon equation in terms of $\hat{\alpha}_{{\rm reg}}$, and then numerically integrate (\ref{Areg-org}), say numerics of $A_{\rm reg}$. To solve the sinh-Gordon equation numerically, we use Mathematica Package NDSolve. Because $p(z)$ is divergent at $z\rightarrow 0,\infty$, we need to introduce  cutoffs close to these singularities. We use the coordinates $x+i y=\log z$ with $x\in [-\Lambda, \Lambda]$ and $y\in [0, 2\pi]$. The cutoff $\Lambda$ is chosen such that the area density near the cutoff is small enough. However, we find that numerical value of the area does not converge but oscillates with a magnitude of order $10^{-4}$ as  $\Lambda$ becomes large. So we cannot get highly accurate results in this approach. 

We have tested the TBA equation numerically for the $(m,n)=(1,1)$ and $(m,n)=(1,2)$ cases and the results are shown in table \ref{tab:m1n1} and \ref{tab:m1n2}, respectively.

\begin{table}[H]
	\centering
	\begin{tabular}{c|c|c}\hline
	$p(z)$& TBA & Numerics \\\hline
	$(z^{-1}+z^{-3})/64$&$0.377385 $&$0.3774$\\
	$(z^{-1}+z^{-2}+z^{-3})/64$ &$0.390659 $ &$0.3906$\\
	$(z^{-1}+\frac{3}{2}z^{-2}+z^{-3})/64$&$0.402791 $&$0.4028$\\
	$(z^{-1}+\frac{12}{5}z^{-2}+z^{-3})/256$&$0.466742 $&$0.4667$\\
	$(z^{-1}+4z^{-2}+z^{-3})/256$ &$0.472178 $ &$0.4722$\\
	$(z^{-1}+6z^{-2}+z^{-3})/512$&$0.419296 $&$0.4193$\\
	$(z^{-1}+10z^{-2}+z^{-3})/1024$&$0.418505$&$0.4185$\\
	\hline
\end{tabular}
	\caption{Comparison of $A_{\rm reg}$ computed by TBA method and the areas obtained by numerically solving the generalized sinh-Gordon equation (\ref{mshG}) and substituting the solution into (\ref{Areg-org}) for some cases with $m=n=1$. To provide a nontrivial test, the coefficients in $p(z)$ are chosen such that the periods $Z_{E_i}$ are not too large and thus the integral terms in the TBA are not small.}
	\label{tab:m1n1}
\end{table}
 
\begin{table}[H]
	\centering
	\begin{tabular}{c|c|c}\hline
	$p(z)$& TBA & Numerics \\\hline
	$ (1+z^{-3})/4$&$0.408174  $&$0.4083$\\
	$ (1+z^{-3})/16$ &$0.503204 $ &$0.5031$\\
	$ (1+z^{-3})/64$&$0.669739  $&$0.6694$\\
	$ (1+z^{-1}+z^{-3})/4$&$0.409190 $&$0.4094$\\
	$(1+z^{-1}+z^{-3})/16$ &$0.498584  $ &$0.4986$\\
	$(1+z^{-1}+z^{-3})/64$&$0.659836 $&$0.6597$\\
	\hline
\end{tabular}
	\caption{Comparison of $A_{\rm reg}$ computed by TBA method and the area obtained from numerical integration for some cases with $m=1$ and $n=2$. }
	\label{tab:m1n2}
\end{table}

\section{Area from the cross ratios}\label{sec:phy-con}

As mentioned in the previous section, the TBA-like equations do not depend on the Wilson lines configuration in an explicit way. To resolve this problem,  we eliminate the central charge $Z_i$ in the TBA equations (\ref{Ieq-x}) by using the cross ratios
\be
\ba\label{Ztocr}
&\log\chi_{i}^{+}=Z_{i}+\bar{Z}_{i}+\int_{\mathbb{R}}\frac{d\theta^{\prime}}{2\pi}\frac{\log{\cal F}_{i}(\theta^{\prime}-\frac{\pi i}{2}+i\phi)}{\cosh(-i\phi-\theta^{\prime})},\\
&\log\chi_{i}^{-}=-iZ_{i}+i\bar{Z}_{i}+\int_{\mathbb{R}}\frac{d\theta^{\prime}}{2\pi}\frac{\log{\cal F}_{i}(\theta^{\prime}-\frac{\pi i}{2}+i\phi)}{\cosh(-i\phi+\frac{\pi i}{2}-\theta^{\prime}-i\epsilon)},
\ea
\ee
where the coordinates ${\cal X}_i$ at $\zeta=1$ and $\zeta=i$ as are denoted as $\chi_i^+$ and $\chi_i^-$ respectively.

 From the asymptotic (\ref{asymptotics}) the cross ratios computed from (\ref{Ztocr}) are positive at least in the limit of large $|Z_E|$.
However, physical values of the cross ratios can be negative.
In Appendix \ref{app:crossratios}, we show a physical configuration of Wilson line where all cross ratios are negative. 
To allow negative cross ratios we expect the asymptotics of an $\mathcal X$-coordinate corresponding to a negative cross ratio  are modified by
\be
{\cal X}^{(0)}_E=-\exp\left( \frac{1}{\zeta}Z_E+\zeta \bar{Z}_E\right) \label{neg-asymptotics}.
\ee
However this asymptotics is not possible if $\alpha$ is analytic except at $z=0,\infty$.
As discussed in \cite{Caetano:2012ac}, if $\alpha \sim \frac{1}{2} \log p\bar p$ at a zero $z_i$, the connection (\ref{eq:connection}) will have a singularity and leads to a monodromy -1 around the zero $z_i$. Therefore the asymptotics (\ref{neg-asymptotics}) corresponds to the case where $\alpha$ has a logarithmic singularity $\alpha \sim \frac{1}{2} \log p\bar p$ near one of the two zeros contained in the quadrilateral $Q_E$ and at another zero $\alpha$ is analytic. The left hand side of integral equations (\ref{Ieq-xA}) should be modified by replacing $\log\mathcal{X}_{i}\rightarrow \log(-\mathcal{X}_{i})$ if  the asymptotics of $\mathcal{X}_{i}$ is given by (\ref{neg-asymptotics}).
Equation (\ref{eq:afin}) should be modified as
\begin{equation}
    A_{\mathrm{fin}}=2A_{\mathrm{reg}}+\frac{\pi}{2}(m+n)-\pi n_s,
\end{equation}
where  $n_s$ is the number of zeros where $\alpha$ takes the singular asymptotics.
In \cite{Caetano:2012ac} the logarithmic singularities at zeros are introduce to have a non-singular world-sheet metric.
The physical interpretation of the singularity here is not clear to us at this point.

To test the correspondence between the asymptotics (\ref{neg-asymptotics}) and logarithmic singularity at a zero
we computed $A_{\rm reg}$  in these two ways for $p(z)=c(\frac{1}{z^3}+\frac{1}{z})$ and $p(z)=c(\frac{1}{z^3}+\frac{1}{z^2}+\frac{1}{z})$  with $\alpha\sim\frac{1}{2} \log p\bar p$ at  $z=i$ and $z=-\frac{1}{2}+\frac{i \sqrt{3}}{2}$ respectively.
The results are summarized in table \ref{tab:Areg-neg-mn=11a}. 
\begin{table}[H]
	\centering
	\begin{tabular}{c|c|c}\hline
	$p(z)$& TBA & Numerics \\\hline
	$1/8(z^{-1}+z^{-3})$&$0.254021 $&$0.2541$\\
	$1/16(z^{-1}+z^{-3})$ &$0.234482$ &$0.2345$\\
	$1/32(z^{-1}+z^{-3})$&$0.197911$&$0.1979$\\
	$1/64(z^{-1}+z^{-3})$&$0.149841$&$0.1498$\\
	$1/8(z^{-1}+z^{-2}+z^{-3})$&$0.225485 $&$0.2255$\\
	$1/16(z^{-1}+z^{-2}+z^{-3})$ &$0.195409$ &$0.1954$\\
	$1/32(z^{-1}+z^{-2}+z^{-3})$&$0.157793$&$0.1578$\\
	$1/64(z^{-1}+z^{-2}+z^{-3})$&$0.116639$&$0.1166$\\
	\hline
\end{tabular}
	\caption{Comparison of $A_{\rm reg}$ computed by TBA method and the area obtained by numerical integration for various $p(z)$.}
	\label{tab:Areg-neg-mn=11a}
\end{table}

Solving equations (\ref{Ztocr}) about the central charge $Z_i^\pm$ and then substituting it to the original TBA equations (\ref{Ieq-xA}), we get
\be\label{crTBA}
\ba
\log\big({\pm\cal X}^-_{i}(\theta+i\phi)\big)=&-i\sinh(\theta+i\phi)\log|\chi_{i}^{+}|-\cosh(\theta+i\phi)\log|\chi_{i}^{-}|\\
&-\int_{\mathbb{R}}\frac{d\theta^{\prime}}{2\pi i}\frac{\sinh(2\theta+2i\phi)}{\sinh(\theta-\theta^{\prime}+i\epsilon)\sinh(2\theta^{\prime}+2i\phi)}\log{\cal F}_{i}(\theta^{\prime}-\frac{\pi i}{2}+i\phi).
\ea
\ee
The choice of signs in the left hand side depend on the the sign of the asymptotics of $\mathcal{X}_{i}$.
To compute areas of worldsheet ending on two periodic  Wilson lines, one solves the integral equations (\ref{crTBA}) with the physical cross ratios as input. Then one can compute the central charges $Z_i$ from (\ref{Ztocr}) and the area $A_{\mathrm{reg}}$ from (\ref{Areg-TBA}). The WKB triangulation depends on the coefficient in $p(z)$ and the choice of $\phi$.
The original TBA equations (\ref{Ieq-xA}) are derived by assuming $p(z)$ is not far away from $ c(z^{-3}+z^{-1})$ and $0<\phi<\pi/2$.
Therefore in the end one need to compute $p(z)$ form the central charges and verify whether $p(z)$ and $\phi$ give the desired  WKB triangulation consistent with  (\ref{crTBA}).

For example, when $m=n=1$, the cross ratios are
\be
\chi_{1}^{\pm}=-\frac{1}{(1+l^{\pm}/q^{\pm})^2},\quad\chi_{2}^{\pm}=-(l^{\pm}/q^{\pm})^2.
\ee
Because the cross ratios are negative, (\ref{crTBA}) takes the form:
\begin{equation}
  \begin{split}\label{eq:TBA11cr}
   &\log ({-\cal X}^-_{1}(\theta+i\phi))=-\cosh(\theta+i\phi)\log\Big(\frac{1}{(1+l^-/q^-)^2}\Big)
   -i\sinh(\theta+i\phi)\log\Big(\frac{1}{(1+l^+/q^+)^2}\Big)\\
&\qquad\qquad\qquad-\int_{\mathbb{R}}\frac{d\theta^{\prime}}{2\pi i}\frac{\sinh(2\theta+2i\phi)}{\sinh(\theta-\theta^{\prime}+i \epsilon)\sinh(2\theta^{\prime}+2i\phi)}\log\Big(\frac{1}{1+\mathcal X_2^-(\theta'+i\phi)}\Big)^2,\\
&\log({-\cal X}^-_{2}(\theta+i\phi))=
-\cosh(\theta+i\phi)\log\Big((\frac{l^-}{q^-})^2\Big)
-i\sinh(\theta+i\phi)\log\Big((\frac{l^+}{q^+})^2 \Big)\\
&\qquad\qquad -\int_{\mathbb{R}}\frac{d\theta^{\prime}}{2\pi i}\frac{\sinh(2\theta+2i\phi)}{\sinh(\theta-\theta^{\prime}+i \epsilon)\sinh(2\theta^{\prime}+2i\phi)}\log\Big(1+\mathcal{X}_1^-(\theta'+i\phi)\big(1+\mathcal{X}_2^-(\theta'+i\phi)\big)^2 \Big)^2.
  \end{split}  
\end{equation}
In order that the integrals in (\ref{eq:TBA11cr}) converge, we require 
\begin{equation}\label{eq:converge}
   \tan \phi \log(l^+/q^+)^2 <\log(l^-/q^-)^2,~~~
   \tan \phi \log(1+l^+/q^+)^2<\log(1+l^-/q^-)^2,
\end{equation}
such that $\mathcal X_{i}(\theta+i\phi)\rightarrow 0$ for $\theta \rightarrow\pm \infty$.
For a given worlsheet, the cross ratios are not uniquely defined because of the equivalence relation $l\simeq l+q$. Using the freedom of choice of $l$, one can always find a value of $\phi\in(0,\pi/2)$ such that the $\mathcal X$-functions decay at large $\theta$.
Then one can numerically solve the integral equations \eqref{eq:TBA11cr} and compute the central charges $Z_i$ from (\ref{Ztocr}) and the area $A_{\mathrm{reg}}$ from (\ref{Areg-TBA}).

\begin{figure}[tbh]
\begin{center}
\resizebox{135mm}{!}{\includegraphics{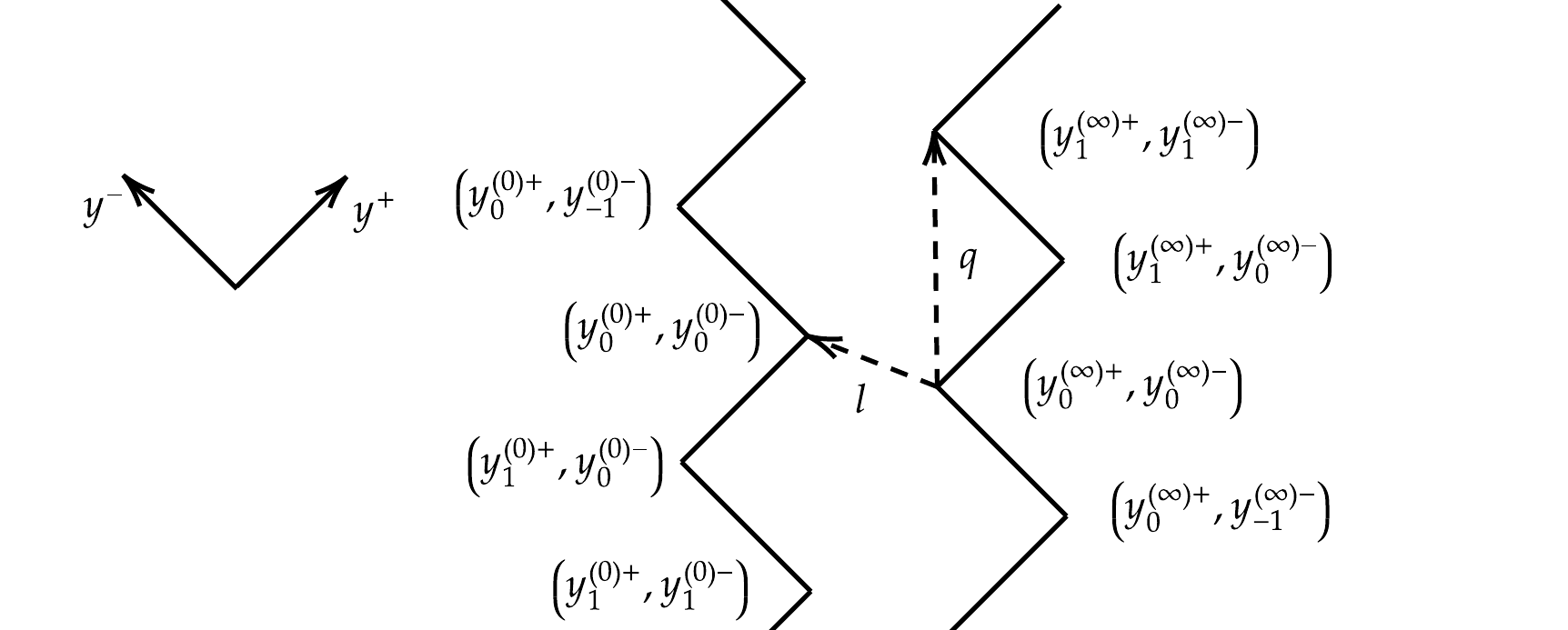}}
\end{center}
  \caption{ The periodic Wilson lines configuration in the case $(m,n)=(1,1)$.} 
\label{fig:ws11}
\end{figure}

\begin{figure}[H]
\begin{center}
\resizebox{90mm}{!}{\includegraphics{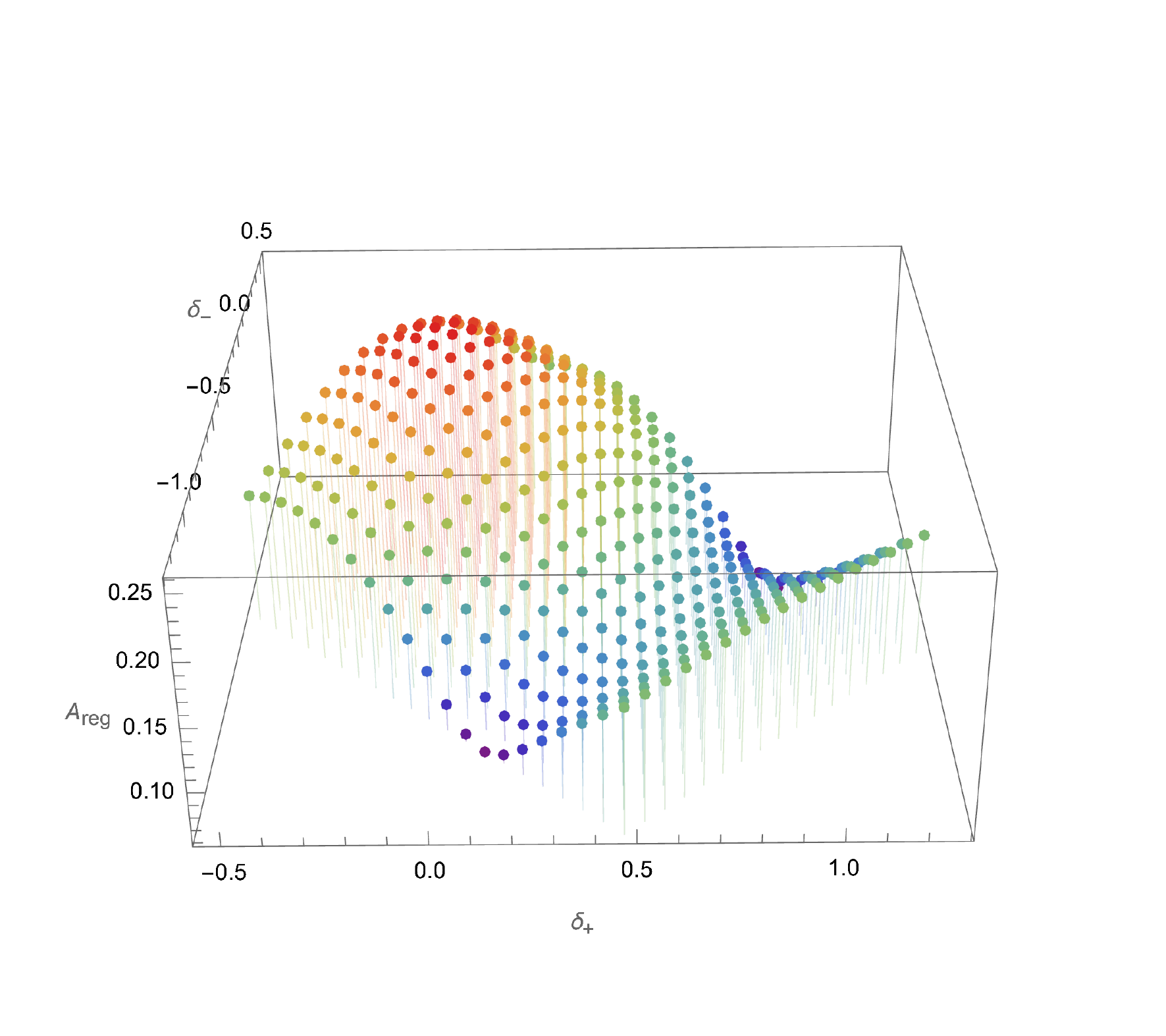}}~~~
\resizebox{10mm}{!}{\includegraphics{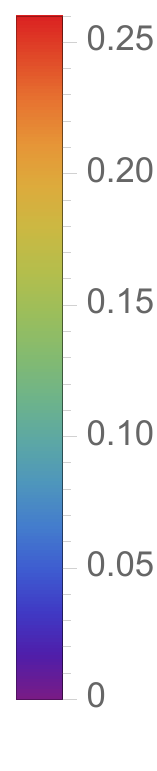}}~~
\end{center}
  \caption{ $A_{\mathrm{reg}}$ as a function of $\delta_{\pm}$ computed by solving (\ref{eq:TBA11cr}) numerically.} 
\label{fig:areg11}
\end{figure}

In general, finding $p(z)$ for given central charges is a difficult task. We consider a special case when $(l^+/q^+,l^-/q^-)\rightarrow(0,-1)$ and thus two Wilson lines almost coincide (see Fig. \ref{fig:ws11}).
In this case some of the logarithms of the  cross ratios diverges and the integrals in (\ref{eq:TBA11cr}) can be negligible. The solution of (\ref{eq:TBA11cr}) can be approximated by its asymptotics and we find:  
\begin{align}
  &A_{\mathrm{reg}}\approx\frac{\pi}{12}
  -\frac{1}{{\sqrt{8\pi }}}\left(\delta_- ^2 \sqrt{-\log \delta_- ^2}
  +\delta_+ ^2 \sqrt{-\log \delta_+ ^2}\right),\\
  &Z_1\approx -\frac{1}{2} i \log \delta_-^2,~~~
Z_2\approx \frac{1}{2}  \log \delta_+^2,~~~\\
&p(z)\approx
\frac{\Gamma \left(\frac{1}{4}\right)^4 \left(\log \delta _-^2+\log \delta _+^2\right){}^2}{4096 \pi ^3}
\left(
\frac{1}{z^3}+\frac{32 \pi ^2 \left(\log \delta _-^2-\log \delta _+^2\right)}{z^2\Gamma \left(\frac{1}{4}\right)^4 \left(\log \delta _-^2+\log \delta _+^2\right)}+\frac{1}{z}
\right)     ,
\end{align}
where both $\delta_+=l^+/p^+$ and $\delta_-=l^-/p^-+1$ are small of the same order in this limit. 
For finite values of $\delta_\pm$, one has to solve (\ref{eq:TBA11cr}) numerically.
The results for some cases when $\delta_+=\pm\delta_-$ are shown in table \ref{tab:crTBA11a} and \ref{tab:crTBA11b}. 
In the case $\delta_+=-\delta_-$ where we separate the two Wilson lines in the transverse direction, $A_{\mathrm{reg}}$ first decreases and then increases after reaching a minimum at $|\delta_\pm|\approx 0.6$ as $|\delta_\pm|$ increases. When two Wilson lines are separated in the longitudinal direction $\delta_-=\delta_+$, we find $A_{\mathrm{reg}}(\delta_\pm=\delta)$ is equal to $A_{\mathrm{reg}}(\delta_\pm=-1-\delta)$, which is consistent with the symmetry of exchanging two Wilson lines. More results are shown in  Fig. \ref{fig:areg11}. We find the $A_{\mathrm{reg}}$ minimizes at  $(\delta_+,\delta_-)=(0,-1)$ where $A_{\mathrm{reg}}\rightarrow 0$. The configuration is equivalent to set $l\rightarrow 0$ and the two Wilson lines in one period become a rectangular Wilson loop, whose minimal area is completely fixed by the dual conformal symmetry\footnote{The dual conformal symmetry completely fixes the 4,5-point scattering amplitudes/Wilson loops, whose amplitudes/minimal area can be expressed by using the BDS conjecture \cite{Bern:2005iz} with a trivial $A_{\rm reg}$. For $n\geq 6$, the area starts to differ from the BDS conjecture \cite{Drummond:2008aq,Bern:2008ap}, namely the $A_{\rm reg}$ will be non-trivial.}.
The area  $A_{\mathrm{reg}}$ reaches the maximum value $\pi/12$ when $\delta_\pm=0$ and two Wilson lines coincide.

It is not obvious the areas are the same for $l$ and $l+q$ from the integral equations  (\ref{eq:TBA11cr}). As a consistency check, we compute the areas for some pairs of equivalent configurations $(\delta_+,\delta_-)$ and $(\delta_++2,\delta_-+2)$. The results are shown in table \ref{tab:lsimlq}.

\begin{table}[H]
	\centering
	\begin{tabular}{c|c|c|c|c}\hline
	$\delta_+=-\delta_-$& $Z_{1,2}$ & $c,c_1$ & TBA & Numerics\\\hline
	$0.3$&$\begin{array}{c}
	     -0.295669 + 1.18518 i  \\
	     -1.18518 + 0.295669 i
	\end{array}$ &$\begin{array}{c}
	     0.030972  \\
	     0.454030 i
	\end{array}$&0.200458&0.2004\\\hline
	$0.4$ &$\begin{array}{c}
	     -0.385816+0.881248 i  \\
	     -0.881248+0.385816 i
	\end{array}$ &$\begin{array}{c}
	     0.0175655  \\
	     0.789943 i
	\end{array}$&0.168922&0.1689\\\hline
	$0.5$&$\begin{array}{c}
	     -0.465244+0.638576 i  \\
	     -0.638576+0.465244 i
	\end{array}$ & $\begin{array}{c}
	    0.00980096  \\
	     1.28713 i
	\end{array}$ &0.143341&0.1433\\\hline
	$0.6$&$\begin{array}{c}
	     -0.530455+0.437494 i  \\
	     -0.437494+0.530455 i
	\end{array}$ &$\begin{array}{c}
	     0.00527798  \\
	     2.03765 i
	\end{array}$&0.130700&0.1307\\\hline
	$0.7$&$\begin{array}{c}
	     -0.581248+0.270192 i  \\
	     -0.270192+0.581248 i
	\end{array}$ &$\begin{array}{c}
	     0.00274072  \\
	     3.20468 i
	\end{array}$&0.133519&0.1335
	\\\hline
		$0.8$&$\begin{array}{c}
	     -0.620828 + 0.131943 i  \\
	     -0.131943 + 0.620828 i
	\end{array}$ &$\begin{array}{c}
	     0.00138794 \\
	     5.07530 i
	\end{array}$&0.148697&0.1487
	\\\hline
\end{tabular}
	\caption{We solve (\ref{eq:TBA11cr}) numerically and compute $Z_i$ and $A_{\rm reg}$ when $\delta_+=-\delta_-$, say ``TBA'' in the table. We choose $\phi=\pi/4$. The number $c$ and $c_1$ are coefficients in $p(z)=c(z^{-3}+c_1z^{-2}+z^{-1})$.
	 We find $c$ is real and $c_i$ is imaginary when $\delta_+=-\delta_-$. The results match the areas (\ref{Areg-org}) obtained by numerical integration, say ``Numerics'' in the table.}
	\label{tab:crTBA11a}
\end{table}
\begin{table}[H]
	\centering
	\begin{tabular}{c|c|c|c|c}\hline
	$\delta_+=\delta_-$& $Z_{1,2}$ & $c$ & TBA & Numerics\\\hline
	$-0.1$&$\begin{array}{c}
	     0.100703+2.30092 i  \\
	     -2.30092+0.100703 i
	\end{array}$ &$0.115028\, -0.010088 i$&$0.252587$&$0.2527$\\\hline
	$-0.2$&$\begin{array}{c}
	     0.20635 +1.60181 i  \\
	     -1.60181+0.20635 i
	\end{array}$ &$0.0549271 -0.0143906 i$&$0.230416$&$0.2304$\\\hline
	$-0.3$ &$\begin{array}{c}
	     0.324384 +1.1854 i  \\
	     -1.1854+0.324384 i
	\end{array}$ &$0.0282984 -0.0167413 i$&$0.201106$&$0.2011$\\\hline
	$-0.4$&$\begin{array}{c}
	    0.466530 +0.883201 i  \\
	     -0.883201+0.466530 i
	\end{array}$ & $0.0122427 -0.0179392 i$ &$0.173421$&$0.1734$\\\hline
	$-0.5$&$\begin{array}{c}
	     0.648457 +0.648457 i  \\
	    -0.648457+0.648457 i
	\end{array}$ &$-0.0183074 i$&$0.161262$&$0.1613$\\\hline
	$-0.6$&$\begin{array}{c}
	    0.883201 + 0.46653 i  \\
	     -0.46653 + 0.883201 i
	\end{array}$ & $-0.0122427-0.0179392 i$ &$0.173421$&$0.1734$\\\hline
	$-0.7$ &$\begin{array}{c}
	     1.1854+0.324384 i  \\
	     -0.324384 +1.1854 i
	\end{array}$ &$-0.0282984 -0.0167413 i$&$0.201106$&$0.2011$\\\hline
\end{tabular}
	\caption{We solve (\ref{eq:TBA11cr}) numerically and compute $Z_i$ and $A_{\rm reg}$ when $\delta_+=\delta_-$. The number $c$ is defined as $p(z)=c(z^{-3}+z^{-1})$. We find the coefficient of $z^{-2}$ vanishes when $\delta_+=\delta_-$. The results match the  areas (\ref{Areg-org}) obtained by numerical integration.}
	\label{tab:crTBA11b}
\end{table}

\begin{table}[H]
	\centering
	\begin{tabular}{c|c|c|c}\hline
	$\delta_+,\delta_-$&$\phi$& $Z_{1,2}$ & $A_{\rm reg}$\\\hline
	$0.5,1.5$&$0.49\pi$&$\begin{array}{c}
	     -0.638582 - 0.465243 i  \\
	     -0.465251 - 0.638577 i
	\end{array}$ &0.143342\\\hline
	$-1.5,-0.5$&$0.01\pi$&$\begin{array}{c}
	     0.638576 +0.465243 i  \\
	    0.465243 +0.638574 i
	\end{array}$ &0.143342\\\hline
		$0.6,1.4$&$0.49\pi$&$\begin{array}{c}
	     -0.881253-0.385815 i  \\
	     -0.385825 - 0.881249  i
	\end{array}$ &0.168923\\\hline
	$-1.4,-0.6$&$0.01\pi$&$\begin{array}{c}
	     0.881248 + 0.385815 i  \\
	    0.385816 + 0.881246 i
	\end{array}$ &0.168922 \\\hline
		$0.7,1.3$&$0.49\pi$&$\begin{array}{c}
	     -0.270198 - 0.581249 i  \\
	     -0.581254 - 0.270193 i
	\end{array}$ &0.133519\\\hline
	$-1.3,-0.7$&$0.01\pi$&$\begin{array}{c}
	     0.270191 + 0.581246 i  \\
	   0.581248 + 0.270192 i
	\end{array}$ &0.133518 \\\hline
\end{tabular}
	\caption{The areas and central charges for some equivalent pairs of configurations computed by solving (\ref{eq:TBA11cr}). In order that (\ref{eq:converge}) is satisfied we need to choose different $\phi$ in each case.}
	\label{tab:lsimlq}
\end{table}

\section{Conclusions and discussion}\label{sec:con-dis}

In this paper,  we have computed the minimal area of the worldsheet ending on two  periodic light-like polygonal Wilson lines at the AdS boundary, which is dual to a cylindrically cut double trace scattering amplitude in the four dimensional ${\cal N}=4$ super Yang-Mills theory. We have presented a boundary condition of the linear problem, which is equivalent to the equation of motion, to produce the two light-like polygonal Wilson lines at the boundary. By using the connection between the linear problem and TBA equations, we have provided an exact method to compute the minimal area ending on the Wilson lines with fixed period $q$ and distance $l$ in the AdS$_3$ subspace. Given the cross ratios as inputs, we have expressed the non-trivial part of the minimal area in terms of the free energy of the TBA system, which matches with the area calculated by using the numerical integration. 

Clearly, there are many open questions raised by our work. Let us mention some of them. 

At the large $l_{\perp}$ limit, we expect some of the moduli parameters of $p(z)$ to be very large, see for example $c_1$ in table \ref{tab:crTBA11a}. In this case, the zeros of $p(z)$ will be separated into two regions far away from each other. Some zeros surround the origin and others are at infinity, which corresponds to the two Wilson lines respectively. Moreover, from the numerical integration of the area, we found the nontrivial contribution is almost coming from the regions around the zeros in this case. One thus can expect the two Wilson lines to decouple in this case, which thus becomes two copies of the form factor case \cite{Maldacena2010}. It would be interesting to see this decomposition analytically.

To allow negative cross ratios, we have modified the boundary condition of $\alpha$ at zeros of polynomial $p(z)$, it would be important to see whether this modification is useful in the case scattering amplitude and provide a physical interpretation to this modification. In this paper, we have focused on the AdS$_3$ subspace for simplicity. It is important to generalize our method to the Wilson lines at the AdS$_5$ boundary to study more general non-planar scattering amplitudes, where one needs to handle the linear problem with a higher rank connection. Some hints on this directions can be found in \cite{Gao:2013dza,Ito:2016qzt,Ito:2021boh,Ito:2021sjo}. It would be interesting to see what happens in various limits of the TBA equations in the AdS$_5$ case \cite{Ito:2018puu,Abl:2021hhb,Lin:2021lqo}, which may shed light on developing the method for the finite coupling.  We hope to address this question in a future publication.
Moreover, our method used to compute the minimal area is quite general. It would be interesting to use it to study the minimal area surface related to the higher order non-planar corrections of the scattering amplitudes or (non-planar) form factor/Wilson lines dual at strong coupling \footnote{See \cite{Lin:2020dyj} for the recent development at the weak coupling.}.

On the other hand, the non-perturbative integrability method turns out to be a very powerful method to compute the planar scattering amplitude/Wilson loop even at the finite coupling region \cite{Alday:2010ku,Basso:2013vsa,Basso:2013aha,Basso:2014koa,Basso:2014nra}, where the Wilson loop is decomposed into square and pentagons. The strong coupling of this approach leads to the Y-system/TBA equations derived from the minimal area surface \cite{Bonini:2015lfr}. Moreover, the scattering amplitude in the colinear limit at strong coupling map into correlators of twist fields in the $O(6)$ sigma model. A surprising consequence of this identification is that an additional exponentially large term due to the sphere $S^5$ will contribute to the scattering amplitude \cite{Basso:2014jfa,Bonini:2018mkg}. It would be interesting to see this contribution in our case. More recently, this approach has been generalized to the form factor case \cite{Sever:2020jjx,Sever:2021nsq,Sever:2021xga}. It would be interesting to explore the case of non-planar scattering amplitudes.

\subsection*{Acknowledgements}
We would like to thank Alfredo Bonini, Davide Fioravanti, Daniele Gregori, Katsushi Ito, Simone Piscaglia, Marco Rossi, Yuji Satoh, Amit Sever, Roberto Tateo, Gang Yang and Hao Zou for useful discussions. We are grateful to Amit Sever and Gang Yang for providing useful comments and suggestions on earlier version of the draft. A part of this work was done when H.O and H.S. were at Nordita and supported by the grant “Exact Results in Gauge and String Theories” from the Knut and Alice Wallenberg foundation.
 H.S. would like to thank Jilin University and Soochow University for their (online) hospitality.

\appendix

\section{More on the $m=n$ case}\label{sec:m=n}
In this appendix, we study the case of $m=n$ in details. We will present a simplified functional relations and TBA equations, whose relation with four dimensional ${\cal N}=2$ SYM will also be mentioned. 

From the functional relations of $m=n$ case (\ref{eq:A-X-rel}), one finds 
\be
\ba
\big({\cal X}_{-k,k}^{[2]}\mathcal{X}_{-k,k}\big)^{-1}=\big(1+A_{-k,k-1}\big)\big(1+A_{-k-1,k}\big)=(1+\mathcal{X}_{-k,k-1})(1+\mathcal{X}_{-k-1,k}).
\ea
\ee
Comparing with $A_{-k,k}$, it is easy to find
\be
A_{-k,k}=\big({\cal X}_{-k,k}^{[2]}\big)^{-1},
\ee
 which can also be checked by using the Pl\"ucker relation. Introducing 
\be
\ba
{\cal Y}_{-k,k}=1/{\cal X}_{-k,k},\quad {\cal Y}_{-k-1,k}={\cal X}_{-k-1,k}^{[-]},
\ea
\ee
we obtain the simplified functional relations for the $m=n$ case:
\be
\ba
{\cal Y}_{-k,k}^{[-]}{\cal Y}_{-k,k}^{[+]}&=\big(1+{\cal Y}_{-k,k-1}\big)\big(1+{\cal Y}_{-k-1,k}),\\
{\cal Y}_{-k-1,k}^{[-]}{\cal Y}_{-k-1,k}^{[+]}&=\big(1+{\cal Y}_{-k,k}\big)\big(1+{\cal Y}_{-k-1,k+1}\big),
\ea
\ee
which is an analogy of the Y-system of the scattering amplitude in \cite{Alday:2010vh}. We thus can derive the TBA equations by following the procedure in the case of scattering amplitude
\be\label{eq:TBAm=n}
\ba
\log{\cal Y}_{-k,k}(\theta+i\phi_{-k,k})=&-|Z_{-k,k}|(e^{-\theta}+e^{\theta})+\int_{\mathbb{R}}\frac{d\theta^{\prime}}{2\pi}\frac{\log\big(1+{\cal Y}_{-k,k-1}(\theta^{\prime}+i\phi_{-k,k-1})\big)}{\cosh(\theta-\theta^{\prime}+i\phi_{-k,k}-i\phi_{-k,k-1})}\\
&+\int_{\mathbb{R}}\frac{d\theta^{\prime}}{2\pi}\frac{\log\big(1+{\cal Y}_{-k-1,k}(\theta^{\prime}+i\phi_{-k-1,k})\big)}{\cosh(\theta-\theta^{\prime}+i\phi_{-k,k}-i\phi_{-k-1,k})},\\
\log{\cal Y}_{-k-1,k}(\theta+i\phi_{-k-1,k})=&-|Z_{-k-1,k}|(e^{-\theta}+e^{\theta})+\int_{\mathbb{R}}\frac{d\theta^{\prime}}{2\pi}\frac{\log\big(1+{\cal Y}_{-k,k}(\theta^{\prime}+i\phi_{-k,k})\big)}{\cosh(\theta-\theta^{\prime}+i\phi_{-k-1,k}-i\phi_{-k,k})}\\&+\int_{\mathbb{R}}\frac{d\theta^{\prime}}{2\pi}\frac{\log\big(1+{\cal Y}_{-k-1,k+1}(\theta^{\prime}+i\phi_{-k-1,k+1})\big)}{\cosh(\theta-\theta^{\prime}+i\phi_{-k-1,k}-i\phi_{-k-1,k+1})},
\ea
\ee
where $\phi$ is defined by
\be
Z_{-k,k}e^{-i\phi_{-k,k}}=|Z_{-k,k}|,\quad -iZ_{-k-1,k}e^{-i\phi_{-k-1,k}}=|Z_{-k-1,k}|.
\ee
The non-trivial part of area can be written by
\be
\ba
 A_{\mathrm{reg}}
=&\frac{\pi}{24}(m+n)+\sum_{2k+1=1}\int_{\mathbb{R}}\frac{d\theta}{2\pi}|Z_{-k,k}|\cosh\theta\log\big(1+{\cal Y}_{-k,k}(\theta+i\phi_{-k,k})\big)\\&+\sum_{2k=2}\int_{\mathbb{R}}\frac{d\theta}{2\pi}|Z_{-k-1,k}|\cosh\theta\log\big(1+{\cal Y}_{-k-1,k}(\theta+i\phi_{-k-1,k})\big),
\ea
\ee
which appears as the free energy of the TBA equations.

By using the new Y-functions, one can express the cross ratios by
\be
\ba
{\cal Y}_{-k,k}(\theta=0)=&1/\chi_{-k,k}^{+},\quad{\cal Y}_{-k,k}(\theta=\frac{\pi i}{2})=1/\chi_{-k,k}^{-},\\
{\cal Y}_{-k-1,k}(\theta=\frac{\pi i}{2})=&\chi_{-k-1,k}^{+},\quad{\cal Y}_{-k-1,k}(\theta=\pi i)=\chi_{-k-1,k}^{-}.
\ea
\ee
We thus are able to rewrite the TBA equations 
\be
\ba
&\log{\cal Y}_{-k,k}(\theta+i\phi_{-k,k})
=-\cosh(\theta+i\phi_{-k,k})\log\chi_{-k,k}^{+}+i\sinh(\theta+i\phi_{-k,k})\log\chi_{-k,k}^{-}\\
&\qquad 
-\int_{\mathbb{R}}\frac{d\theta^{\prime}}{2\pi}\frac{\sinh\big(2(\theta+i\phi_{-k,k})\big)\log\big(1+{\cal Y}_{-k,k-1}(\theta^{\prime}+i\phi_{-k,k-1})\big)}{\cosh(\theta-\theta^{\prime}+i\phi_{-k,k}-i\phi_{-k,k-1})\sinh\big(2(\theta^{\prime}+i\phi_{-k,k-1})\big)}\\
&
\qquad
-\int_{\mathbb{R}}\frac{d\theta^{\prime}}{2\pi}\frac{\sinh\big(2(\theta+i\phi_{-k,k})\big)\log\big(1+{\cal Y}_{-k-1,k}(\theta^{\prime}+i\phi_{-k-1,k})\big)}{\sinh\big(2(\theta^{\prime}+i\phi_{-k-1,k})\big)\cosh\big(\theta-\theta^{\prime}+-i\phi_{-k-1,k}+i\phi_{-k,k}\big)},\\
&\log{\cal Y}_{-k-1,k}(\theta+i\phi_{-k-1,k})
=-\cosh(\theta+i\phi_{-k-1,k})\log\chi_{-k-1,k}^{-}-i\sinh(\theta+i\phi_{-k-1,k})\log\chi_{-k-1,k}^{+}\\
&\qquad-\int_{\mathbb{R}}\frac{d\theta^{\prime}}{2\pi}\frac{\sinh\big(2(\theta+i\phi_{-k-1,k})\big)\log\big(1+{\cal Y}_{-k,k}(\theta^{\prime}+i\phi_{-k,k})\big)}{\cosh(\theta-\theta^{\prime}+i\phi_{-k-1,k}-i\phi_{-k,k})\sinh\big(2(\theta^{\prime}+i\phi_{-k,k}\big)}\\
&\qquad-\int_{\mathbb{R}}\frac{d\theta^{\prime}}{2\pi}\frac{\sinh\big(2(x+i\phi_{-k-1,k})\big)\log\big(1+{\cal Y}_{-k-1,k+1}(\theta^{\prime}+i\phi_{-k-1,k+1})\big)}{\sinh\big(2(\theta^{\prime}+i\phi_{-k-1,k+1})\big)\cosh(\theta-\theta^{\prime}+i\phi_{-k-1,k}-i\phi_{-k-1,k+1})}.
\ea
\ee

It is worth to note that the TBA equations \eqref{eq:TBAm=n} with $m=n=1$ coincide with the ones of the ${\cal N}=2$ pure $SU(2)$ SYM in \cite{GrassiGuMarino,FioravantiGregori:2019}. The reason is because, when $m=n=1$, the Riemann surface \eqref{eq:Rie-sur} appears as the Seiberg-Witten curve of the pure $SU(2)$ SYM, which is thus based on the similar mathematical structure. Keeping the relation with Seiberg-Witten theory in mind, we find a hidden relations of the TBA equations \eqref{eq:TBAm=n}.
Let us parametrize the curve of the $m=n=1$ case by $p(z)=\frac{1}{z^3}+\frac{2u}{z^2}+\frac{1}{z}$, whose central charge $Z_1$ and $Z_2$ are given by
\be
\ba
Z_{1}(u)=i\big(\Pi_{A}(u)+\Pi_{B}(u)\big),\quad Z_{1}(u)=-i\Pi_{B}(u),
\ea
\ee
where 
\be
\ba
&\Pi_{A}(u)=8\sqrt{2+2u}{\rm E}(\frac{4}{2+2u}),\\
&\Pi_{B}(u)=8i\sqrt{2+2u}\Big({\rm K}(\frac{2u-2}{2+2u})-{\rm E}(\frac{2u-2}{2+2u})\Big),
\ea
\ee
for $0<u<1$.
It is easy to find $Z_{2}(-u)=iZ_{1}(u)$. The TBA equations for $m=n=1$ case thus can be written as
\be
\ba
\log{\cal Y}_{0,0}(\theta+i\frac{\pi}{2},u)&=|Z_{0,0}(u)|(e^{-\theta}+e^{\theta})-\int_{\mathbb{R}}\frac{d\theta^{\prime}}{\pi}\frac{\log\big(1+\frac{1}{{\cal Y}_{-1,0}(\theta^{\prime}+\frac{\pi}{2},u)}\big)}{\cosh(\theta-\theta^{\prime})},\\
\log{\cal Y}_{-1,0}(\theta+i\frac{\pi}{2},u)&=|Z_{0,0}(-u)|(e^{-\theta}+e^{\theta})-\int_{\mathbb{R}}\frac{d\theta^{\prime}}{\pi}\frac{\log\big(1+\frac{1}{{\cal Y}_{0,0}(\theta^{\prime}+i\frac{\pi}{2},u)}\big)}{\cosh(\theta-\theta^{\prime})}.
\ea
\ee
 We thus can find
\be
{\cal Y}_{-1,0}(\theta+i\frac{\pi}{2},-u)={\cal Y}_{0,0}(\theta+i\frac{\pi}{2},u).
\ee

\section{A simple example of physical configuration} \label{app:crossratios}

For the $m=n$ case, we consider the following configuration of coordinates on the boundary:
\begin{equation}
	y_k^{(\infty)\pm}=\frac{q^{\pm}}{n} k,~~~y_{-k}^{(0)\pm}=\frac{q^{\pm}}{n} k+l^{\pm},
\end{equation}
where $k=1,2,...,n$.
The cross ratios can be computed as
\begin{equation}
\begin{split}
	-\frac{\left(y^{(\infty)\pm}_{k-1}-y^{(\infty)\pm}_k\right) \left(y^{(0)\pm}_{-k-1}-y^{(0)\pm}_{-k}\right)}{\left(y^{(\infty)\pm}_k-y^{(0)\pm}_{-k-1}\right) \left(y^{(0)\pm}_{-k}-y^{(\infty)\pm}_{k-1}\right)}&=
	-\frac{(q^\pm)^{ 2}}{(q^\pm+nl^\pm)^2},
\\
-\frac{\left(y^{(0)\pm}_{1-k}-y^{(\infty)\pm}_{k-1}\right) \left(y^{(\infty)\pm}_k-y^{(0)\pm}_{-k}\right)}{\left(y^{(\infty)\pm}_{k-1}-y^{(\infty)\pm}_k\right) \left(y^{(0)\pm}_{-k}-y^{(0)\pm}_{1-k}\right)}	
&=-\frac{n^2(l^\pm)^{ 2}}{(q^\pm)^{ 2}}.
\end{split}
\end{equation}
Therefore the $\mathcal{X}$-functions at $\zeta=1,i$ are related to $l$ and $q$ as
\begin{equation}
	\mathcal{X}_{-k,k}(\zeta=1,i)=-\frac{(q^\pm)^{2}}{(q^\pm+nl^\pm)^2},~~~
	\mathcal{X}_{-k-1,k}(\zeta=1,i)=-\frac{(nl^\pm)^{2}}{(q^\pm)^2}.
\end{equation}
We expect the solution to the functional relations takes the form
\begin{equation}
	{\cal X}_{-k,k}={\cal X}_{-0,0},~~~{\cal X}_{-k-1,k}={\cal X}_{-1,0},
\end{equation}
for all $k$ and thus reduce to the $m=n=1$ case.

\providecommand{\href}[2]{#2}\begingroup\raggedright\endgroup


\begin{thebibliography}{10}

\bibitem{Maldacena:1997re}
J.~M. Maldacena, \emph{{The Large N limit of superconformal field theories and
  supergravity}}, \href{http://dx.doi.org/10.1023/A:1026654312961}{\emph{Adv.
  Theor. Math. Phys.} {\bf 2} (1998) 231--252},
  [\href{http://arxiv.org/abs/hep-th/9711200}{{\tt hep-th/9711200}}].

\bibitem{Beisert:2010jr}
N.~Beisert et~al., \emph{{Review of AdS/CFT Integrability: An Overview}},
  \href{http://dx.doi.org/10.1007/s11005-011-0529-2}{\emph{Lett. Math. Phys.}
  {\bf 99} (2012) 3--32}, [\href{http://arxiv.org/abs/1012.3982}{{\tt
  1012.3982}}].

\bibitem{Alday:2007hr}
L.~F. Alday and J.~M. Maldacena, \emph{{Gluon scattering amplitudes at strong
  coupling}},
  \href{http://dx.doi.org/10.1088/1126-6708/2007/06/064}{\emph{JHEP} {\bf 06}
  (2007) 064}, [\href{http://arxiv.org/abs/0705.0303}{{\tt 0705.0303}}].

\bibitem{Maldacena:1998im}
J.~M. Maldacena, \emph{{Wilson loops in large N field theories}},
  \href{http://dx.doi.org/10.1103/PhysRevLett.80.4859}{\emph{Phys. Rev. Lett.}
  {\bf 80} (1998) 4859--4862}, [\href{http://arxiv.org/abs/hep-th/9803002}{{\tt
  hep-th/9803002}}].

\bibitem{Rey:1998ik}
S.-J. Rey and J.-T. Yee, \emph{{Macroscopic strings as heavy quarks in large N
  gauge theory and anti-de Sitter supergravity}},
  \href{http://dx.doi.org/10.1007/s100520100799}{\emph{Eur. Phys. J. C} {\bf
  22} (2001) 379--394}, [\href{http://arxiv.org/abs/hep-th/9803001}{{\tt
  hep-th/9803001}}].

\bibitem{Pohlmeyer:1975nb}
K.~Pohlmeyer, \emph{{Integrable Hamiltonian Systems and Interactions Through
  Quadratic Constraints}},
  \href{http://dx.doi.org/10.1007/BF01609119}{\emph{Commun. Math. Phys.} {\bf
  46} (1976) 207--221}.

\bibitem{Alday:2009yn}
L.~F. Alday and J.~Maldacena, \emph{{Null polygonal Wilson loops and minimal
  surfaces in Anti-de-Sitter space}},
  \href{http://dx.doi.org/10.1088/1126-6708/2009/11/082}{\emph{JHEP} {\bf 11}
  (2009) 082}, [\href{http://arxiv.org/abs/0904.0663}{{\tt 0904.0663}}].

\bibitem{Alday:2009dv}
L.~F. Alday, D.~Gaiotto and J.~Maldacena, \emph{{Thermodynamic Bubble Ansatz}},
  \href{http://dx.doi.org/10.1007/JHEP09(2011)032}{\emph{JHEP} {\bf 09} (2011)
  032}, [\href{http://arxiv.org/abs/0911.4708}{{\tt 0911.4708}}].

\bibitem{Burrington:2009bh}
B.~A. Burrington and P.~Gao, \emph{{Minimal surfaces in AdS space and
  Integrable systems}},
  \href{http://dx.doi.org/10.1007/JHEP04(2010)060}{\emph{JHEP} {\bf 04} (2010)
  060}, [\href{http://arxiv.org/abs/0911.4551}{{\tt 0911.4551}}].

\bibitem{Burrington:2011eh}
B.~A. Burrington, \emph{{General Leznov-Savelev solutions for Pohlmeyer reduced
  AdS$_5$ minimal surfaces}},
  \href{http://dx.doi.org/10.1007/JHEP09(2011)002}{\emph{JHEP} {\bf 09} (2011)
  002}, [\href{http://arxiv.org/abs/1105.3227}{{\tt 1105.3227}}].

\bibitem{GaiottoMooreNeitzke:2008}
D.~Gaiotto, G.~W. Moore and A.~Neitzke, \emph{{Four-dimensional wall-crossing
  via three-dimensional field theory}},
  \href{http://dx.doi.org/10.1007/s00220-010-1071-2}{\emph{Commun. Math. Phys.}
  {\bf 299} (2010) 163--224}, [\href{http://arxiv.org/abs/0807.4723}{{\tt
  0807.4723}}].

\bibitem{GaiottoMooreNeitzke:2009}
D.~Gaiotto, G.~W. Moore and A.~Neitzke, \emph{{Wall-crossing, Hitchin Systems,
  and the WKB Approximation}},  \href{http://arxiv.org/abs/0907.3987}{{\tt
  0907.3987}}.

\bibitem{Dorey:1998pt}
P.~Dorey and R.~Tateo, \emph{{Anharmonic oscillators, the thermodynamic Bethe
  ansatz, and nonlinear integral equations}},
  \href{http://dx.doi.org/10.1088/0305-4470/32/38/102}{\emph{J. Phys. A} {\bf
  32} (1999) L419--L425}, [\href{http://arxiv.org/abs/hep-th/9812211}{{\tt
  hep-th/9812211}}].

\bibitem{Bazhanov:1998wj}
V.~V. Bazhanov, S.~L. Lukyanov and A.~B. Zamolodchikov, \emph{{Spectral
  determinants for Schrodinger equation and Q operators of conformal field
  theory}}, \href{http://dx.doi.org/10.1023/A:1004838616921}{\emph{J. Statist.
  Phys.} {\bf 102} (2001) 567--576},
  [\href{http://arxiv.org/abs/hep-th/9812247}{{\tt hep-th/9812247}}].

\bibitem{Lukyanov:2010rn}
S.~L. Lukyanov and A.~B. Zamolodchikov, \emph{{Quantum Sine(h)-Gordon Model and
  Classical Integrable Equations}},
  \href{http://dx.doi.org/10.1007/JHEP07(2010)008}{\emph{JHEP} {\bf 07} (2010)
  008}, [\href{http://arxiv.org/abs/1003.5333}{{\tt 1003.5333}}].

\bibitem{Alday:2010vh}
L.~F. Alday, J.~Maldacena, A.~Sever and P.~Vieira, \emph{{Y-system for
  Scattering Amplitudes}},
  \href{http://dx.doi.org/10.1088/1751-8113/43/48/485401}{\emph{J. Phys. A}
  {\bf 43} (2010) 485401}, [\href{http://arxiv.org/abs/1002.2459}{{\tt
  1002.2459}}].

\bibitem{Hatsuda:2010cc}
Y.~Hatsuda, K.~Ito, K.~Sakai and Y.~Satoh, \emph{{Thermodynamic Bethe Ansatz
  Equations for Minimal Surfaces in $AdS_{3}$}},
  \href{http://dx.doi.org/10.1007/JHEP04(2010)108}{\emph{JHEP} {\bf 04} (2010)
  108}, [\href{http://arxiv.org/abs/1002.2941}{{\tt 1002.2941}}].

\bibitem{Hatsuda:2010vr}
Y.~Hatsuda, K.~Ito, K.~Sakai and Y.~Satoh, \emph{{Six-point gluon scattering
  amplitudes from $\mathbb{Z}_{4}$-symmetric integrable model}},
  \href{http://dx.doi.org/10.1007/JHEP09(2010)064}{\emph{JHEP} {\bf 09} (2010)
  064}, [\href{http://arxiv.org/abs/1005.4487}{{\tt 1005.4487}}].

\bibitem{Ito:2018eon}
K.~Ito, M.~Mari\~no and H.~Shu, \emph{{TBA equations and resurgent Quantum
  Mechanics}}, \href{http://dx.doi.org/10.1007/JHEP01(2019)228}{\emph{JHEP}
  {\bf 01} (2019) 228}, [\href{http://arxiv.org/abs/1811.04812}{{\tt
  1811.04812}}].

\bibitem{Ito:2019jio}
K.~Ito and H.~Shu, \emph{{TBA equations for the Schr\"odinger equation with a
  regular singularity}},
  \href{http://dx.doi.org/10.1088/1751-8121/ab96ee}{\emph{J. Phys. A} {\bf 53}
  (2020) 335201}, [\href{http://arxiv.org/abs/1910.09406}{{\tt 1910.09406}}].

\bibitem{Emery:2020qqu}
Y.~Emery, \emph{{TBA equations and quantization conditions}},
  \href{http://dx.doi.org/10.1007/JHEP07(2021)171}{\emph{JHEP} {\bf 07} (2021)
  171}, [\href{http://arxiv.org/abs/2008.13680}{{\tt 2008.13680}}].

\bibitem{Fioravanti:2020udo}
D.~Fioravanti, M.~Rossi and H.~Shu, \emph{{$QQ$-system and non-linear integral
  equations for scattering amplitudes at strong coupling}},
  \href{http://dx.doi.org/10.1007/JHEP12(2020)086}{\emph{JHEP} {\bf 12} (2020)
  086}, [\href{http://arxiv.org/abs/2004.10722}{{\tt 2004.10722}}].

\bibitem{Alday:2007he}
L.~F. Alday and J.~Maldacena, \emph{{Comments on gluon scattering amplitudes
  via AdS/CFT}},
  \href{http://dx.doi.org/10.1088/1126-6708/2007/11/068}{\emph{JHEP} {\bf 11}
  (2007) 068}, [\href{http://arxiv.org/abs/0710.1060}{{\tt 0710.1060}}].

\bibitem{Maldacena2010}
J.~Maldacena and A.~Zhiboedov, \emph{{Form factors at strong coupling via a
  Y-system}}, \href{http://dx.doi.org/10.1007/JHEP11(2010)104}{\emph{JHEP11}
  (2010) 104}.

\bibitem{Gao:2013dza}
Z.~Gao and G.~Yang, \emph{{Y-system for form factors at strong coupling in
  $AdS_5$ and with multi-operator insertions in $AdS_3$}},
  \href{http://dx.doi.org/10.1007/JHEP06(2013)105}{\emph{JHEP} {\bf 06} (2013)
  105}, [\href{http://arxiv.org/abs/1303.2668}{{\tt 1303.2668}}].

\bibitem{Ben-Israel:2018ckc}
R.~Ben-Israel, A.~G. Tumanov and A.~Sever, \emph{{Scattering amplitudes
  \textemdash{} Wilson loops duality for the first non-planar correction}},
  \href{http://dx.doi.org/10.1007/JHEP08(2018)122}{\emph{JHEP} {\bf 08} (2018)
  122}, [\href{http://arxiv.org/abs/1802.09395}{{\tt 1802.09395}}].

\bibitem{Janik:2011bd}
R.~A. Janik and A.~Wereszczynski, \emph{{Correlation functions of three heavy
  operators: The AdS contribution}},
  \href{http://dx.doi.org/10.1007/JHEP12(2011)095}{\emph{JHEP} {\bf 12} (2011)
  095}, [\href{http://arxiv.org/abs/1109.6262}{{\tt 1109.6262}}].

\bibitem{Kazama:2011cp}
Y.~Kazama and S.~Komatsu, \emph{{On holographic three point functions for GKP
  strings from integrability}},
  \href{http://dx.doi.org/10.1007/JHEP01(2012)110}{\emph{JHEP} {\bf 01} (2012)
  110}, [\href{http://arxiv.org/abs/1110.3949}{{\tt 1110.3949}}].

\bibitem{Caetano:2012ac}
J.~Caetano and J.~Toledo, \emph{{$\chi$-systems for correlation functions}},
  \href{http://dx.doi.org/10.1007/JHEP01(2019)050}{\emph{JHEP} {\bf 01} (2019)
  050}, [\href{http://arxiv.org/abs/1208.4548}{{\tt 1208.4548}}].

\bibitem{Hitchin:1986vp}
N.~J. Hitchin, \emph{{The Selfduality equations on a Riemann surface}},
  \href{http://dx.doi.org/10.1112/plms/s3-55.1.59}{\emph{Proc. Lond. Math.
  Soc.} {\bf 55} (1987) 59--131}.

\bibitem{Iwaki2014}
K.~Iwaki and T.~Nakanishi, \emph{{Exact WKB analysis and cluster algebras}},
  \href{http://dx.doi.org/10.1088/1751-8113/47/47/474009}{\emph{Journal of
  Physics A: Mathematical and Theoretical} {\bf 47} (2014) },
  [\href{http://arxiv.org/abs/1401.7094}{{\tt 1401.7094}}].

\bibitem{Bern:2005iz}
Z.~Bern, L.~J. Dixon and V.~A. Smirnov, \emph{{Iteration of planar amplitudes
  in maximally supersymmetric Yang-Mills theory at three loops and beyond}},
  \href{http://dx.doi.org/10.1103/PhysRevD.72.085001}{\emph{Phys. Rev. D} {\bf
  72} (2005) 085001}, [\href{http://arxiv.org/abs/hep-th/0505205}{{\tt
  hep-th/0505205}}].

\bibitem{Drummond:2008aq}
J.~M. Drummond, J.~Henn, G.~P. Korchemsky and E.~Sokatchev, \emph{{Hexagon
  Wilson loop = six-gluon MHV amplitude}},
  \href{http://dx.doi.org/10.1016/j.nuclphysb.2009.02.015}{\emph{Nucl. Phys. B}
  {\bf 815} (2009) 142--173}, [\href{http://arxiv.org/abs/0803.1466}{{\tt
  0803.1466}}].

\bibitem{Bern:2008ap}
Z.~Bern, L.~J. Dixon, D.~A. Kosower, R.~Roiban, M.~Spradlin, C.~Vergu et~al.,
  \emph{{The Two-Loop Six-Gluon MHV Amplitude in Maximally Supersymmetric
  Yang-Mills Theory}},
  \href{http://dx.doi.org/10.1103/PhysRevD.78.045007}{\emph{Phys. Rev. D} {\bf
  78} (2008) 045007}, [\href{http://arxiv.org/abs/0803.1465}{{\tt 0803.1465}}].

\bibitem{Ito:2016qzt}
K.~Ito and H.~Shu, \emph{{ODE/IM correspondence for modified $B_2^{(1)}$ affine
  Toda field equation}},
  \href{http://dx.doi.org/10.1016/j.nuclphysb.2017.01.009}{\emph{Nucl. Phys. B}
  {\bf 916} (2017) 414--429}, [\href{http://arxiv.org/abs/1605.04668}{{\tt
  1605.04668}}].

\bibitem{Ito:2021boh}
K.~Ito, T.~Kondo, K.~Kuroda and H.~Shu, \emph{{WKB periods for higher order ODE
  and TBA equations}},
  \href{http://dx.doi.org/10.1007/JHEP10(2021)167}{\emph{JHEP} {\bf 10} (2021)
  167}, [\href{http://arxiv.org/abs/2104.13680}{{\tt 2104.13680}}].

\bibitem{Ito:2021sjo}
K.~Ito, T.~Kondo and H.~Shu, \emph{{Wall-crossing of TBA equations and WKB
  periods for the third order ODE}},
  \href{http://arxiv.org/abs/2111.11047}{{\tt 2111.11047}}.

\bibitem{Ito:2018puu}
K.~Ito, Y.~Satoh and J.~Suzuki, \emph{{MHV amplitudes at strong coupling and
  linearized TBA equations}},
  \href{http://dx.doi.org/10.1007/JHEP08(2018)002}{\emph{JHEP} {\bf 08} (2018)
  002}, [\href{http://arxiv.org/abs/1805.07556}{{\tt 1805.07556}}].

\bibitem{Abl:2021hhb}
T.~Abl and M.~Sprenger, \emph{{Exploring Reggeon bound states in
  strongly-coupled $ \mathcal{N} $ = 4 super Yang-Mills}},
  \href{http://dx.doi.org/10.1007/JHEP01(2022)021}{\emph{JHEP} {\bf 01} (2022)
  021}, [\href{http://arxiv.org/abs/2108.02302}{{\tt 2108.02302}}].

\bibitem{Lin:2021lqo}
G.~Lin, G.~Yang and S.~Zhang, \emph{{Color-Kinematics Duality and Dual
  Conformal Symmetry for A Four-loop Form Factor in N=4 SYM}},
  \href{http://arxiv.org/abs/2112.09123}{{\tt 2112.09123}}.

\bibitem{Lin:2020dyj}
G.~Lin and G.~Yang, \emph{{Non-planar form factors of generic local operators
  via on-shell unitarity and color-kinematics duality}},
  \href{http://dx.doi.org/10.1007/JHEP04(2021)176}{\emph{JHEP} {\bf 04} (2021)
  176}, [\href{http://arxiv.org/abs/2011.06540}{{\tt 2011.06540}}].

\bibitem{Alday:2010ku}
L.~F. Alday, D.~Gaiotto, J.~Maldacena, A.~Sever and P.~Vieira, \emph{{An
  Operator Product Expansion for Polygonal null Wilson Loops}},
  \href{http://dx.doi.org/10.1007/JHEP04(2011)088}{\emph{JHEP} {\bf 04} (2011)
  088}, [\href{http://arxiv.org/abs/1006.2788}{{\tt 1006.2788}}].

\bibitem{Basso:2013vsa}
B.~Basso, A.~Sever and P.~Vieira, \emph{{Spacetime and Flux Tube S-Matrices at
  Finite Coupling for N=4 Supersymmetric Yang-Mills Theory}},
  \href{http://dx.doi.org/10.1103/PhysRevLett.111.091602}{\emph{Phys. Rev.
  Lett.} {\bf 111} (2013) 091602}, [\href{http://arxiv.org/abs/1303.1396}{{\tt
  1303.1396}}].

\bibitem{Basso:2013aha}
B.~Basso, A.~Sever and P.~Vieira, \emph{{Space-time S-matrix and Flux tube
  S-matrix II. Extracting and Matching Data}},
  \href{http://dx.doi.org/10.1007/JHEP01(2014)008}{\emph{JHEP} {\bf 01} (2014)
  008}, [\href{http://arxiv.org/abs/1306.2058}{{\tt 1306.2058}}].

\bibitem{Basso:2014koa}
B.~Basso, A.~Sever and P.~Vieira, \emph{{Space-time S-matrix and Flux-tube
  S-matrix III. The two-particle contributions}},
  \href{http://dx.doi.org/10.1007/JHEP08(2014)085}{\emph{JHEP} {\bf 08} (2014)
  085}, [\href{http://arxiv.org/abs/1402.3307}{{\tt 1402.3307}}].

\bibitem{Basso:2014nra}
B.~Basso, A.~Sever and P.~Vieira, \emph{{Space-time S-matrix and Flux-tube
  S-matrix IV. Gluons and Fusion}},
  \href{http://dx.doi.org/10.1007/JHEP09(2014)149}{\emph{JHEP} {\bf 09} (2014)
  149}, [\href{http://arxiv.org/abs/1407.1736}{{\tt 1407.1736}}].

\bibitem{Bonini:2015lfr}
A.~Bonini, D.~Fioravanti, S.~Piscaglia and M.~Rossi, \emph{{Strong Wilson
  polygons from the lodge of free and bound mesons}},
  \href{http://dx.doi.org/10.1007/JHEP04(2016)029}{\emph{JHEP} {\bf 04} (2016)
  029}, [\href{http://arxiv.org/abs/1511.05851}{{\tt 1511.05851}}].

\bibitem{Basso:2014jfa}
B.~Basso, A.~Sever and P.~Vieira, \emph{{Collinear Limit of Scattering
  Amplitudes at Strong Coupling}},
  \href{http://dx.doi.org/10.1103/PhysRevLett.113.261604}{\emph{Phys. Rev.
  Lett.} {\bf 113} (2014) 261604}, [\href{http://arxiv.org/abs/1405.6350}{{\tt
  1405.6350}}].

\bibitem{Bonini:2018mkg}
A.~Bonini, D.~Fioravanti, S.~Piscaglia and M.~Rossi, \emph{{Fermions and
  scalars in $\mathcal{N} = 4$ Wilson loops at strong coupling and beyond}},
  \href{http://dx.doi.org/10.1016/j.nuclphysb.2019.114644}{\emph{Nucl. Phys. B}
  {\bf 944} (2019) 114644}, [\href{http://arxiv.org/abs/1807.09743}{{\tt
  1807.09743}}].

\bibitem{Sever:2020jjx}
A.~Sever, A.~G. Tumanov and M.~Wilhelm, \emph{{Operator Product Expansion for
  Form Factors}},
  \href{http://dx.doi.org/10.1103/PhysRevLett.126.031602}{\emph{Phys. Rev.
  Lett.} {\bf 126} (2021) 031602}, [\href{http://arxiv.org/abs/2009.11297}{{\tt
  2009.11297}}].

\bibitem{Sever:2021nsq}
A.~Sever, A.~G. Tumanov and M.~Wilhelm, \emph{{An Operator Product Expansion
  for Form Factors II. Born level}},
  \href{http://dx.doi.org/10.1007/JHEP10(2021)071}{\emph{JHEP} {\bf 10} (2021)
  071}, [\href{http://arxiv.org/abs/2105.13367}{{\tt 2105.13367}}].

\bibitem{Sever:2021xga}
A.~Sever, A.~G. Tumanov and M.~Wilhelm, \emph{{An Operator Product Expansion
  for Form Factors III. Finite Coupling and Multi-Particle Contributions}},
  \href{http://arxiv.org/abs/2112.10569}{{\tt 2112.10569}}.

\bibitem{GrassiGuMarino}
A.~Grassi, J.~Gu and M.~Marino, \emph{{Non-perturbative approaches to the
  quantum Seiberg--Witten curve}},  \href{http://arxiv.org/abs/1908.07065}{{\tt
  1908.07065}}.

\bibitem{FioravantiGregori:2019}
D.~Fioravanti and D.~Gregori, \emph{{Integrability and cycles of deformed
  ${\cal N}=2$ gauge theory}},
  \href{http://dx.doi.org/10.1016/j.physletb.2020.135376}{\emph{Phys. Lett. B}
  {\bf 804} (2020) 135376}, [\href{http://arxiv.org/abs/1908.08030}{{\tt
  1908.08030}}].

\end{thebibliography}
\end{document}